\documentclass[aps,prd,reprint,onecolumn,nofootinbib]{revtex4-2}
\usepackage{graphicx}
\usepackage{xcolor}
\usepackage{hyperref}
\usepackage{amsmath}
\usepackage{graphicx}
\usepackage{slashed}
\usepackage{multirow}
\usepackage[caption=false]{subfig}
\usepackage{pdfsync}%Inverse search
\graphicspath{{figures/}}
\graphicspath{{JaxoFig_MuonToEGamma/}}
\DeclareGraphicsExtensions{.png,.pdf}

\begin{document}
	
	\title{Lepton-flavor violating decays induced by Lorentz violation in the Yukawa sector of the Standard Model Extension}

	\author{J. Montaño-Domínguez ${}^{a}$}
	\author{F. Ram\'irez-Zavaleta ${}^{b}$}
	\author{E. S. Tututi ${}^{b}$}
	\author{O. Vázquez-Hernández ${}^{a}$}

	\affiliation{${}^{a}$ SECIHTI, Av. Insurgentes Sur 1582, Col. Cr\'edito Constructor, Alc. Benito Ju\'arez, 03940, CDMX, M\'exico.}
	\affiliation{${}^{b}$ Facultad de Ciencias F\'isico Matem\'aticas, Universidad Michoacana de San Nicol\'as de Hidalgo, Av. Francisco J. M\'ugica s/n, C.P. 58060, Morelia, Michoac\'an, M\'exico.}

	\date{February 19, 2026}

		\begin{abstract}
		Tree-level lepton-flavor-violating decays induced by Lorentz-violating effects within the Yukawa sector of the Standard Model Extension are studied. These new physics effects are parameterized by the $(Y_{f})_{\mu\nu}^{AB}$ tensor, with $\mu$ and $\nu$ denoting Lorentz indices and $A$, $B$ being indices in the flavor space. Since this tensor is antisymmetric under the interchange of Lorentz indices, in analogy with the electromagnetic tensor $F_{\mu\nu}$, $(Y_{f})_{\mu\nu}^{AB}$ can be expressed in terms of six components associated with two complex three-vectors denoted by $\mathbf{e}_l^{AB}$ and $\mathbf{b}_l^{AB}$. On the assumption that these three-vectors are pure real or pure imaginary and mutually orthogonal, constraints on their magnitudes via experimental bounds on lepton-flavor-violating processes $\mathrm{Br}(\l_B\rightarrow \gamma l_A)$ and $\mathrm{Br}(\l_B\rightarrow l_A l_C \bar{l}_C)$ are estimated. Thus, the $\l_B\rightarrow \gamma l_A$ decay provides the following upper bounds: $\lvert \mathbf{e}_l^{\mu\tau} \lvert < 1.51\times 10^{-11}$ , $\lvert \mathbf{e}_l^{e\tau} \lvert < 1.34 \times 10^{-11}$, $\lvert \mathbf{e}_l^{e\mu} \lvert < 3.65 \times 10^{-18}$, $\lvert \mathbf{b}_l^{\mu\tau} \lvert < 1.95\times 10^{-11}$, $\lvert \mathbf{b}_l^{e\tau} \lvert <1.73\times 10^{-11}$, $\lvert \mathbf{b}_l^{e\mu} \lvert <  4.71\times 10^{-18}$. Conversely, by assuming that the Lorentz-violating parameters are purely real, for the $\l_B\rightarrow l_A l_C \bar{l}_C$ process it is found that $\lvert \mathbf{e}_l^{\mu\tau}\lvert< 3.05 \times 10^{-12}$  and $\lvert \mathbf{b}_l^{\mu\tau}\lvert< 4.31 \times 10^{-12}$. These results offer more restrictive bounds than those previously reported in the literature.
	\end{abstract}

	\maketitle

	\section{Introduction}
The Standard Model (SM) of particle physics provides the most accurate description of the strong and electroweak interactions \cite{Glashow,Salam,Weinberg,Englert1964et,Higgs1964pj,Guralnik1964eu,ATLAS2012yve,CMS2012qbp}, where its Lagrangian is built upon the principles of quantum mechanics, special relativity and gauge invariance. Nevertheless, several investigations suggest that, at fundamental scales, such as the Planck scale, Lorentz symmetry could be explicitly or spontaneously broken  \cite{KosteleckyStringT1,KosteleckyStringT2,KosteleckyStringT3,Gambini,Carroll,Alfaro,Froggatt,Burgess,KosteleckyT4,Jackiw,Arkani,Horava,Pospelov}. The technological limitations in reaching such energy scales motivate the search for signals of Lorentz symmetry breaking in high-sensitivity experiments at  low energies, such as the SM scale~\cite{Bluhm1997,Bluhm1998bv,Dehmelt1999jh,Crivellin2022idw,HurtadoSilva2023pdl,Ahuatzin2010da}. In this regard, low-energy effects of Lorentz violation (LV) can be parameterized in a model-independent framework known as the Standard Model Extension (SME) \cite{ColladaySME,KosteleckySME}. This model is an effective field theory that encompasses the SM along with general relativity and is structured by means of the sum of all independent effective  operators of the form
\begin{eqnarray}\label{OpLV}
	\mathcal{T}^{\mu\nu\lambda\dots}\mathcal{O}_{\mu\nu\lambda\dots}\, ,
\end{eqnarray}
where $\mathcal{T}^{\mu\nu\lambda\dots}$ denotes constant Lorentz tensors and $\mathcal{O}_{\mu\nu\lambda\dots}$ represents CPT-even or CPT-odd gauge invariant operators built on SM fields \cite{ColladaySME,KosteleckySME}. The SME Lagrangian is manifestly invariant under observer Lorentz transformations, also known as passive transformations. Nevertheless, under particle Lorentz transformations, the components of the $\mathcal{T}^{\mu\nu\lambda\dots}$ tensor remain invariants, whilst the $\mathcal{O}_{\mu\nu\lambda\dots}$ ones transforms covariantly breaking Lorentz invariance. In its minimal version, the SME contains only renormalizable terms, i.e., those with mass dimension less  or equal than four. However, at higher energies, nonrenormalizable terms must be considered in order to preserve causality and stability of the theory \cite{KosteleckyStab,Kostelecky2009zp}. The Lorentz-violating tensors $\mathcal{T}^{\mu\nu\lambda\dots}$ can be interpreted as static background relic fields, in this sense, they do not represent new degrees of freedom, but they act on the SM fields as couplings that could lead to preferential space-time directions. Accordingly, since Lorentz invariance is broken under particle transformations, their effects can be studied by varying the spatial orientation of the experimental setup. In this way, various experiments have been developed to search for  LV sources, or for setting bounds on the components of the $\mathcal{T}^{\mu\nu\lambda\dots}$ tensor \cite{Bluhm1997,Kostelecky1999,Bluhm1998bv,Dehmelt1999jh,Bluhm1999dx,Hughes2001yk,Bluhm2001rw,Hou2003zz,Heckel2006ww,Altschul2007kr,Heckel2008hw,Kostelecky2002hh}. A wide variety of studies have assessed the impact of LV, for instance, on quantum electrodynamics interactions up to the one-loop level \cite{Mariz2016ooa,deBrito2016zav,Brito2020eiy,Kostelecky2009zp}. Moreover, studies on effects of LV in electroweak interactions \cite{Iltan2003nq,Iltan2003we,CastroMedina2015zua,HernandezJuarez2018dkx,AhuatziAvendano2020ppm,Gomes2019twi,Aranda2013cva} or in the extended quantum chromodynamics sector \cite{Colladay2007aj,Vieira2016gxp} have been addressed. Special attention has been given to the impact of LV on physical observables such as anomalous dipole moments \cite{AhuatziAvendano2020ppm,Aghababaei2017bei,Crivellin2022idw,MontanoDominguez2022ytp,MontanoDominguez2021dml,HurtadoSilva2023pdl}, neutrino charge radius \cite{FloresOrea2025rtv}, LV running couplings \cite{Colladay2007aj,HernandezJuarez2018dkx, Toscano2023lii}, decay rates and several scattering processes \cite{Iltan2003nq,Iltan2003we,CastroMedina2015zua,deBrito2016zav,Vieira2016gxp,Gomes2019twi,Charneski2012py,Colladay2002bi,OConnor2025pbs}.\\

One of the most interesting features provided by the SME is that it can generate charged flavor-violating transitions \cite{ColladaySME}. On this matter, the neutrino oscillations \cite{SuperKamiokande1998kpq} reveals the presence of lepton-flavor violation, which is of fundamental relevance in particle physics. In contrast, the SM has been unable to explain this phenomenon since it was constructed under the assumption of massless neutrinos, in consequence, lepton-flavor violation (LFV) interactions are forbidden at tree level. Nonetheless, experimental observations have established upper bounds on branching ratios (BR) of various lepton-flavor violating processes such as $l_B \rightarrow l_A \gamma$ or $l_B\rightarrow l_A \bar{l}_C l_C$, with $l_{A,B,C}=e\, , \mu\, , \tau$; their numerical values are given in Table \ref{ExpBounds} \cite{PDG}. Thus, the topic of flavor violation promoted by charged leptons results interesting in order to search for signals of new physics effects as it occurs in the SME.
	\begin{table}[th!]
		\caption{Experimental bounds on lepton-flavor violation for the two and three-body decay processes $l_A\rightarrow l_B\gamma$ and $l_A\rightarrow l_B l_C \bar{l}_C$, respectively.}\label{ExpBounds}
	\centering
	\begin{tabular}{ccc}
		\hline\hline
		\multicolumn{1}{c}{2-body}                         & \multicolumn{1}{c}{} & \multicolumn{1}{c}{3-body}                          \\ \hline\hline
		\multirow{5}{*}{$\mathrm{Br}(\tau^{-} \rightarrow  \mu^{-}\gamma)<4.2\times 10^{-8}$}                       &                       & \multirow{3}{*}{$\mathrm{Br}(\tau^{-} \rightarrow  \mu^{-}\mu^{+}\mu^{-})<1.9\times 10^{-8}$}                        \\
		&                       &                                                      \\
		&                       &                                                      \\
		&                       & \multirow{3}{*}{$\mathrm{Br}(\tau^{-} \rightarrow  e^{-}e^{+}e^{-})<2.7\times 10^{-8}$}                         \\
		&                       &                                                      \\
		\multirow{5}{*}{$\mathrm{Br}(\tau^{-} \rightarrow  e^{-}\gamma)<3.3\times 10^{-8}$}                        &                       &                                                      \\
		&                       & \multirow{3}{*}{$\mathrm{Br}(\mu^{-} \rightarrow  e^{-}e^{+}e^{-})<1.0\times 10^{-12}$}                        \\
		&                       &                                                      \\
		&                       &                                                      \\
		&                       & \multirow{3}{*}{$\mathrm{Br}(\tau^{-} \rightarrow  e^{-}\mu^{+}\mu^{-})<2.7\times 10^{-8}$}                      \\
		\multicolumn{1}{l}{\multirow{5}{*}{$\mathrm{Br}(\mu^{-} \rightarrow e^{-}\gamma)<3.1\times 10^{-13}$}} & \multicolumn{1}{l}{} &                                                      \\
		\multicolumn{1}{l}{}                               & \multicolumn{1}{l}{} &                                                      \\
		\multicolumn{1}{l}{}                               & \multicolumn{1}{l}{} & \multicolumn{1}{l}{\multirow{3}{*}{$\mathrm{Br}(\tau^{-} \rightarrow  \mu^{-}e^{+}e^{-})<1.8\times 10^{-8}$}} \\
		\multicolumn{1}{l}{}                               & \multicolumn{1}{l}{} & \multicolumn{1}{l}{}                                \\
		\multicolumn{1}{l}{}                               & \multicolumn{1}{l}{} & \multicolumn{1}{l}{}                                \\ \hline\hline
	\end{tabular}
\end{table}

Several extensions of the SM contain interactions that can induce LFV processes \cite{Lee1977tib,Bilenky1977du,Marciano1977wx,Buchmuller1985jz,Campbell2003wp,Dassinger2007ru,Giunti2007ry,MEG2016leq,NovalesSanchez2016sng,Teixeira2016ecr,HernandezTome2018fbq,Barrie2022ake}. For example, the minimally extended Standard Model \cite{Giunti2007ry}, in which three right-handed neutrinos are incorporated, predicts a value for $\mathrm{BR}(\mu\rightarrow e\gamma)$ of the order of $10^{-54}$ \cite{Calibbi2017uvl}; also provides branching ratios for the 3-body decays given in Table \ref{ExpBounds} of the order of $\sim 10^{-55}-10^{-56}$. This implies that these processes are extremely suppressed and till now far from being detectable. The effective field theory addresses LFV via dimension-five and dimension-six operators, promisingly, at new-physics energy scales ($\Lambda$) of unities of TeV up to $10^{4}$ TeV \cite{Buchmuller1985jz,Teixeira2016ecr}.	

On the other hand, the responsible part of the  SME for LFV couplings is the Minimal Extended Yukawa (MEY) sector, where LV effects are parametrized by the complex 2-tensor $(Y_f)_{\mu\nu}^{AB}$, with $\mu$, $\nu$ denoting space-time components and $A$, $B$ labeling flavor indices; more details on these parameters will be presented in the next section. To date, some bounds on the maxima of the values of such parameters were derived in Refs.~\cite{Iltan2003nq,AhuatziAvendano2020ppm}. In the former \cite{Iltan2003nq}, tree-level LV effects on the $H_0\rightarrow f^{-}f^{+}$ decay, for $m_{H_0}=100 \, \mathrm{GeV}$, were studied, yielding, for any of the diagonal terms ($A=B$), $10^{-20}<\lvert Y_f \lvert <10^{-17}$, being $\lvert Y_f \lvert^{2} \equiv (Y_f)_{\alpha\beta}^{AA}(Y_f)^{\alpha\beta}{}^{AA}$. In the latter, one-loop contributions of the MEY sector to the lepton magnetic and electric-dipole moments $a_{l}$ and $d_{l}$, respectively, with $l=e,\mu,\tau$, were determined. In that work, only contributions that are independent of the 4-momentum of the particles, i.e., those proportional to scalar-quadratic terms like $(Y_l)_{\alpha\beta}^{AB}(Y_l)^{\alpha\beta}{}^{AB}$, were considered. For the off-diagonal terms ($A\neq B$), the authors established bounds for the case  $A=\mu$, $B=\tau$, where the most stringent bound corresponds to $\lvert(V_l)_{\alpha\beta}^{\mu\tau}(V_l)^{\mu\tau}{}^{\alpha\beta}\lvert< 10^{-15}$, with $(V_l)_{\alpha\beta}^{AB}=1/2[(Y_l)_{\alpha\beta}^{AB}+(Y_l)_{\alpha\beta}^{BA*}]$; meanwhile for the diagonal case $A=B=e$, their most restrictive bound is $\frac{1}{2}\lvert\epsilon^{\alpha\beta\lambda\rho}(V_l)_{\alpha\beta}^{ee}(V_l)_{\lambda\rho}^{ee}\lvert< 10^{-28}$.

In light of previous considerations, LFV decays offer an interesting scenario to study LV effects since their stringent experimental bounds could lead to strong constraints on LV parameters. In the present work, we calculate the impact of LV effects through LFV tree-level decays $l_B \rightarrow l_A \gamma$ and $l_B\rightarrow l_A \bar{l}_C l_C$ (with $l_{A,B,C}=e\, , \mu\, , \tau$), which are induced by the MEY sector. Our manuscript is organized as follows. Section \ref{SecII} presents the MEY Lagrangian. Section \ref{SecIII} is devoted to the calculation of the two and three-body decay rates and the estimation for bounds on LV parameters. Since the amplitudes involve several products of various Dirac matrices, which could induce dipolar terms, we also calculate LV second order corrections to the 3-point vertex  $l_A l_B\gamma$ function ($A\neq B$) in order to investigate if there exist tree-level contributions to transition magnetic or electric-dipole moments. Finally, in Section \ref{SecIV} the conclusions are presented.

		\section{The Yukawa extended sector}\label{SecII}
\begin{figure}[h]
		\includegraphics[trim= -30mm 220mm 0mm 20mm, scale=0.65,clip]{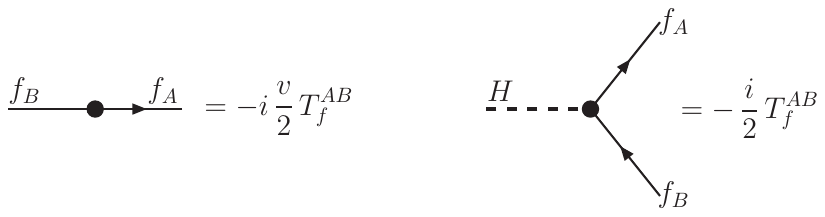}
		\caption{Feynman rules derived from the MEY sector. On the left, the two-point  $f_B f_A$ vertex function. On the right, the $H f_A f_B$ three vertex.}\label{Frules}
	\end{figure}

The Yukawa sector of the SME is given by the Lagrangian \cite{ColladaySME}
	\begin{eqnarray}\label{mYSMEGaugeBasis}
		\mathcal{L}^{\mathrm{mSME}}_{\mathrm{Y}}=\mathcal{L}^{\mathrm{SM}}_{\mathrm{Y}}-\frac{1}{2}(H_{L})_{\mu\nu}^{AB}\bar{L}_{A}\Phi \sigma^{\mu\nu} R_{B}-\frac{1}{2}(H_{U})_{\mu\nu}^{AB}\bar{Q}_{A}\Phi \sigma^{\mu\nu} U_{B}-\frac{1}{2}(H_{D})_{\mu\nu}^{AB}\bar{Q}_{A}\tilde{\Phi} \sigma^{\mu\nu} D_{B}+\mathrm{h.c.},
	\end{eqnarray}
where $Q_A$ and $L_A$ are the quark and lepton doublets of $SU_L(2)$, respectively, while $R_{B}$ is a charged lepton singlet of $SU_L(2)$, $U_{B}$ ($D_{B}$) is an up-type (down-type) quark singlet of $SU_L(2)$, and $\tilde{\Phi}=i\sigma^{2}\Phi$, with $\Phi$ being the SM Higgs doublet. LV effects are induced by the  $(H_{F})_{\mu\nu}^{AB}$ components, with $F=L,U,D$; $\mu$, $\nu$ represent space-time components, whereas $A$, $B$ symbolize either lepton or quark flavor indices. It should be noted that this 2-rank tensor is antisymmetric under the interchange of Lorentz indices, i.e.  $(H_{F})_{\mu\nu}^{AB}=-(H_{F})_{\nu\mu}^{AB}$. After spontaneous symmetry breaking in the unitary gauge, in the mass-eigenstates basis, Eq.~(\ref{mYSMEGaugeBasis}) takes the form
	\begin{eqnarray}
		\mathcal{L}^{\mathrm{mSME}}_{\mathrm{Y}}=-\sum_{f=l,u,d}\bar{f}_A\left\{\left(m_{f_A}+\frac{g}{2 m_W}H\right)\delta^{AB} +\frac{1}{2}(v+H) \Big[\big(Y_f\big)^{AB}_{\mu\nu}P_R +\big(Y_f\big)^{BA*}_{\mu\nu}P_L\Big]\sigma^{\mu\nu}\right\}f_B.
	\end{eqnarray}		
Here, $(Y_f)_{\mu\nu}^{AB}=\sqrt{2}\, V^{f\, \dagger}_{L}(H_f)^{AB}_{\mu\nu}V^{f}_{R}$, being $V^{f}_{L,R}$ the components of a unitary matrix, which diagonalize $\mathcal{L}^{\mathrm{mSME}}_{\mathrm{Y}}$ into the fermion-mass basis; $v$ refers to the vacuum expectation value. Notice that the antisymmetry property in the Lorentz indices for $(H_{F})_{\mu\nu}^{AB}$ is preserved by $(Y_f)_{\mu\nu}^{AB}$. The $P_{L,R}$ correspond to the projectors $\frac{1}{2}(1 \mp \gamma_5)$.  Since $\gamma_5$ will give rise to Levy-Civita tensors, when these are fully contracted with the Lorentz components of the form $(Y_f)^{AB}_{\mu\nu}$, then it will generate the dual tensor
	\begin{eqnarray}
		(\tilde{Y}_{f})^{AB}_{\mu \nu}\equiv \frac{1}{2}\epsilon_{\mu\nu\alpha\beta}(Y_{f})^{AB\,\alpha\beta}.
	\end{eqnarray}
On the other hand, in analogy with the relation between the electric and magnetic 3-vectors, $\mathbf{E}$ and $\mathbf{B}$, respectively, derived from the electromagnetic tensor $F_{\mu\nu}$, two background 3-vectors $\mathbf{e}^{AB}_{f}$ and $\mathbf{b}^{AB}_{f}$, can be introduced as \cite{MontanoDominguez2022ytp}
	\begin{eqnarray}
		(Y_f)^{AB}_{0 i} = (\mathbf{e}^{AB}_{f})_{i},&  (Y_f)^{AB}_{i j} = +\epsilon^{ijk} (\mathbf{b}^{AB}_{f})^{k},\label{ebDef1}\\
		(\tilde{Y}_f)^{AB}_{0 i} = (\mathbf{b}^{AB}_{f})_{i},&	(\tilde{Y}_f)^{AB}_{i j} = -\epsilon^{ijk} (\mathbf{e}^{AB}_{f})^{k}. \label{ebDef2}
	\end{eqnarray}
For off diagonal terms ($A\neq B$), the $\mathcal{L}^{\mathrm{mSME}}_{\mathrm{Y}}$ Lagrangian generates lepton and quark flavor-violating transitions through the 2-point $f_A f_B$ vertex function and from the $H\bar{f}_A f_B$ coupling, at tree level. The respective Feynman rules are given in Fig.~\ref{Frules}.

\noindent where the notation used in Fig.~\ref{Frules} is explicitly given as
	\begin{eqnarray}
		T^{AB}_{f}= \Big[(\mathcal{V}^{AB}_f)_{\mu\nu}+(\mathcal{A}^{AB}_f)_{\mu\nu}\gamma_5\Big]\sigma^{\mu\nu}.
	\end{eqnarray}
	with
	\begin{eqnarray}
		(\mathcal{V}^{AB}_f)_{\mu\nu}&\equiv& \frac{1}{2}\Big[\big(Y_f\big)^{AB}_{\mu\nu} +\big(Y_f\big)^{BA*}_{\mu\nu} \Big],\\
		(\mathcal{A}^{AB}_f)_{\mu\nu}&\equiv& \frac{1}{2}\Big[\big(Y_f\big)^{AB}_{\mu\nu} -\big(Y_f\big)^{BA*}_{\mu\nu} \Big].
	\end{eqnarray}
	However, under the assumption that LV coefficients are symmetric under the interchange of flavor indices it is found that
	\begin{eqnarray}
		(\mathcal{V}^{AB}_f)_{\mu\nu}&\equiv&\,\,\mathrm{Re}\Big[\big(Y_f\big)^{AB}_{\mu\nu}\Big],\\
		(\mathcal{A}^{AB}_f)_{\mu\nu}&\equiv&i \,\mathrm{Im}\Big[\big(Y_f\big)^{AB}_{\mu\nu}\Big].
	\end{eqnarray}
	Hereafter, we will work under this consideration. Besides, from Eqs.~(\ref{ebDef1}) and (\ref{ebDef2}), it can be seen that
	\begin{eqnarray}
		(\mathcal{V}^{AB}_f)_{0i} = \mathrm{Re}\Big[\mathbf{e}^{AB}_{f}\Big]_{i},&  	(\mathcal{V}^{AB}_f)_{ij} = \epsilon^{ijk} \mathrm{Re}\Big[\mathbf{b}^{AB}_{f}\Big]^{k}, \\
		(\mathcal{A}^{AB}_f)_{0i} = i\, \mathrm{Im}\Big[\mathbf{e}^{AB}_{f}\Big]_{i},&  	(\mathcal{A}^{AB}_f)_{ij} = i\, \epsilon^{ijk} \mathrm{Im}\Big[\mathbf{b}^{AB}_{f}\Big]^{k}.
	\end{eqnarray}
In the next section, for simplicity, we will only explore two scenarios: (a) pure real parameters, when $(\mathcal{A}^{AB}_f)_{\mu\nu}=0$ and (b) pure imaginary parameters, for $(\mathcal{V}^{AB}_f)_{\mu\nu}=0$.
	
\section{The calculation}\label{SecIII}
\subsection{Two-body $f_B \rightarrow f_{A}\gamma$ decay}
\begin{figure}[h]
		\includegraphics[trim= -10mm 222mm 0mm 25mm, scale=0.65,clip]{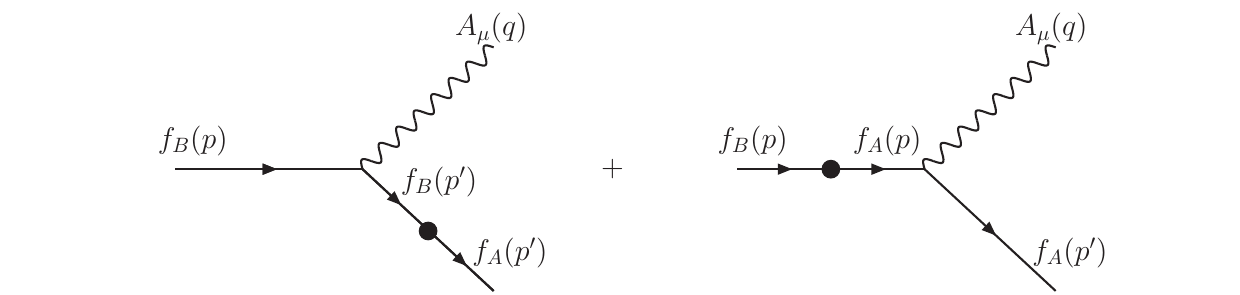}
		\caption{First-order LV corrections to the $f_{B}f_{A}\gamma$ vertex.}\label{BToAG}
	\end{figure}

The first-order LV contributions of the MYE sector to the $f_{B} f_{A}\gamma$ vertex function are described by the Feynman diagrams in Fig. \ref{BToAG}.
The associated amplitude is given by
	\begin{eqnarray}\label{GBToAG}
		\Gamma^{O(1)f_B f_A\gamma}_{\mu}= i\, e\, Q_{f}\ \frac{v}{2}\, \bar{u}^{s^{\prime}}(p^\prime,m_A)\left[ T^{AB}_{f}\left(\frac{\slashed{p}^{\prime}+m_B}{p^{\prime \, 2}-m_B^2}\right)\gamma_{\mu}+\gamma_{\mu}\left(\frac{\slashed{p}+m_A}{p^{2}-m_A^2}\right)T^{AB}_{f}\right]u^{s}(p,m_B),
	\end{eqnarray}
where $Q_f$ is the electric charge of the fermion in consideration. After some Dirac algebraic manipulations, this amplitude can be rewritten as follows
\begin{eqnarray}\label{GBToAG2}
		\Gamma^{O(1)f_B f_A\gamma}_{\mu}=\displaystyle - \frac{2\,  e\, Q_{f} \,v}{(m_B^2-m_A^2)}\, \bar{u}^{s^{\prime}}(p^\prime,m_A)\Bigg\{\gamma^{\alpha}\Big[ (m_A+m_B)(\mathcal{V}^{AB}_f)_{\mu\alpha}-(m_B-m_A)(\mathcal{A}^{AB}_f)_{\mu\alpha}\gamma_5\Big] \nonumber\\
		\qquad\qquad\quad\quad- P^{\alpha}\Big[(\mathcal{V}^{AB}_f)_{\mu\alpha} +(\mathcal{A}^{AB}_f)_{\mu\alpha}\gamma_5 \Big]+i\,  \sigma_{\mu}{}^{\alpha}q^{\beta}\Big[(\mathcal{V}^{AB}_f)_{\alpha\, \beta} +(\mathcal{A}^{AB}_f)_{\alpha\, \beta}\gamma_5\Big] \Bigg\}u^{s}(p,m_B),
	\end{eqnarray}
where $P=p+p^{\prime}$ and $q=p-p^{\prime}$. Since the amplitude satisfies the Ward identity $q^{\mu} \Gamma^{O(1)f_B f_A\gamma}_{\mu}=0$, it is gauge invariant.

From Eq. (\ref{GBToAG2}), it can be observed that there are no contributions to vector or axial currents proportional to $\gamma_{\mu}$ or $\gamma_{\mu}\gamma_{5}$, respectively, nor contributions to the transition-magnetic or  transition-electric dipole moments.

\begin{figure}[ht!]
		\includegraphics[trim= -130mm 155mm 0mm 25mm, scale=0.4,clip]{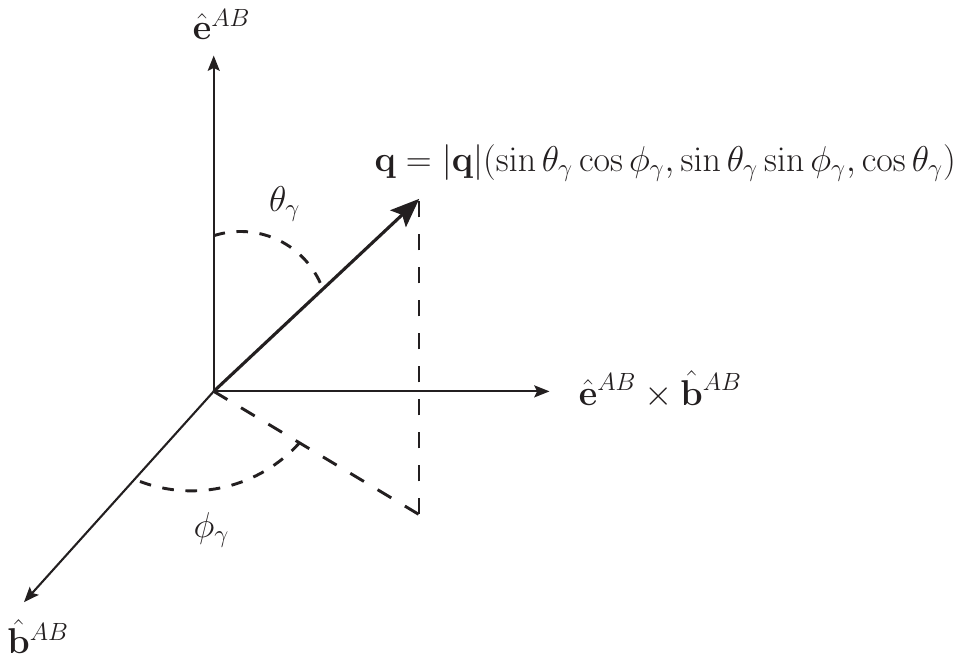}
		\caption{Reference frame conformed by the unitary vectors along the real or imaginary parts of the $\mathbf{e}^{AB}_{f}$, $\mathbf{b}^{AB}_{f}$, and $\mathbf{e}^{AB}_{f}\times\mathbf{b}^{AB}_{f}$ vectors. The $\theta_\gamma$ and $\phi_\gamma$ correspond to the polar and azimuthal angles of the three-momentum of the final photon, respectively.}\label{SR2B}
	\end{figure}

The invariant amplitude for the $f_B(p) \rightarrow f_A(p^{\prime})\gamma(q)$ decay can be determined from Eq. (\ref{GBToAG2}) through
	\begin{eqnarray}
		\mathcal{M}(f_B\rightarrow f_A\gamma)= \Gamma^{O(1)f_B f_A\gamma}_{\mu}\epsilon^{*\mu}(q).
	\end{eqnarray}
Because we are not considering flavor conservation ($A= B$), it will not induce modifications to the energy-momentum dispersion relations of the fermionic fields. Consequently, there are no additional restrictions in the phase-space integration for the decay rates and the average over initial spines by summing final spines remains unchanged; the photon polarization states also remain unmodified \cite{deBrito2016zav,Iltan2003nq,Iltan2003we,CastroMedina2015zua,Gomes2019twi,Aranda2013cva,Vieira2016gxp,Charneski2012py,Colladay2002bi,OConnor2025pbs}. Accordingly, the mean square amplitude can be written as
	\begin{align}\label{MSQ2b}
		\lvert\overline{\mathcal{M}}(f_B\rightarrow f_A\gamma)\lvert^2 = \displaystyle\frac{64\pi\alpha Q_f^2 v^2}{\left(m_B^2-m_A^2\right)^2}\Bigg\{\Bigg[\frac{\left(m_B^2-m_A^2\right)^2}{4}\left(\mathcal{V}^{AB}_{f}\right)^2+\frac{m_B^2+m_A^2}{2}\left(\mathcal{V}^{AB}_f\right)_{\lambda \, q}\left(\mathcal{V}^{AB}_f\right)^{\lambda \, q}\quad \quad\qquad  \nonumber \\ -(m_B^2-m_A^2)(\mathcal{V}^{AB}_f)_{\lambda \, P}(\mathcal{V}^{AB}_{f})^{\lambda \, q}+\frac{\left(\mathcal{V}^{AB}_f\right)_{P \, q}\left(\mathcal{V}^{AB}_f\right)_{P \, q}}{2} -(\mathcal{V}\rightarrow \mathcal{A}) \Bigg]\quad \quad \qquad\nonumber\\
		+\,\, i \,\,\Bigg[\left(m_B-m_A\right)^2\big(\mathcal{V}^{AB}_f\big)_{\lambda \, q}\big(\tilde{\mathcal{A}}^{AB}_f\big)^{\lambda \, q}-\left(m_B+m_A\right)^2\big(\mathcal{A}^{AB}_f\big)_{\lambda \, q}\big(\tilde{\mathcal{V}}^{AB}_f\big)^{\lambda \, q}  \qquad \qquad  \nonumber\\
		-(m_B^2-m_A^2)\left( \big(\mathcal{V}^{AB}_f\big)_{\lambda \, q}\big(\tilde{\mathcal{A}}^{AB}_f\big)^{\lambda \, P} -\left(\mathcal{A}^{AB}_f\right)_{\lambda \, q}\big(\tilde{\mathcal{V}}^{AB}_f\big)^{\lambda \, P}\right)\qquad \quad \quad \quad \qquad \nonumber\\
		+\frac{1}{2}\epsilon^{\rho\tau\alpha\beta} P_{\alpha}q_{\beta}\Big((m_B^2-m_A^2)\left(\mathcal{A}^{AB}_f\right)_{\lambda \, \rho}\left(\mathcal{V}^{AB}_f\right)^{\lambda}{}_{\tau} +\left(\mathcal{A}^{AB}_f\right)_{\rho \, q} \left(\mathcal{V}^{AB}_f\right)_{\tau \, P} \qquad\quad \nonumber \\
		+\left(\mathcal{V}^{AB}_f\right)_{\rho \, q} \left(\mathcal{A}^{AB}_f\right)_{\tau \, P}\Big)\Bigg]\Bigg\}.\qquad \qquad\qquad\qquad\qquad \qquad\qquad\qquad\qquad \qquad\quad\,
	\end{align}
Here, $\big(\mathcal{K}^{AB}_{f}\big)^2=\big(\mathcal{K}^{AB}_{f}\big)_{\lambda\rho}\big(\mathcal{K}^{AB}_{f}\big)^{\lambda\rho}$, $\big(\mathcal{K}^{AB}_f\big)_{\lambda \, r}=r^{\rho}\big(\mathcal{K}^{AB}_f\big)_{\lambda \, \rho}$, and $\big(\tilde{\mathcal{K}}^{AB}_f\big)_{\lambda \, r}=\frac{1}{2}r_{\rho}\epsilon_{\alpha\beta\lambda\rho} \big(\mathcal{K}^{AB}_f\big)^{\alpha\beta}$, with $r=P,q$; $\big(\mathcal{K}^{AB}_f\big)_{P \, q}=P^{\lambda}q^{\rho}\big(\mathcal{K}^{AB}_f\big)_{\lambda\rho}$, where $\mathcal{K}=\mathcal{A},\mathcal{V}$. The contractions between LV tensors and four-momentum components, $P$ and $q$, can be expressed in terms of dot and cross products between the 3-vectors $\mathbf{P}$, $\mathbf{q}$ with the  LV vectors $\mathbf{e}^{AB}_{f}$ and $\mathbf{b}^{AB}_{f}$; these expressions are detailed in Appendix \ref{AppendixA}. Thereby, the decay rate of the $f_B\rightarrow f_A \gamma$ process can be written as
	\begin{eqnarray}
		d\Gamma^{}(f_{B} \rightarrow f_A \gamma)=\frac{1}{2m_B}\left(\frac{1}{2E_{\mathbf{p}^{\prime}}}\frac{d^{3}\mathbf{p}^{\prime}}{(2\pi)^3}\right)\left(\frac{1}{2E_{\mathbf{q}}}\frac{d^{3}\mathbf{q}}{(2\pi)^3}\right)(2\pi)^4 \delta^{(4)}(p-p^{\prime}-q)\lvert\overline{\mathcal{M}}(f_B\rightarrow f_A\gamma)\lvert^2.
	\end{eqnarray}
The integration over the  volume element $d^{3}\mathbf{p}^{\prime}$ is carried out at the rest frame of the decaying particle, where $\mathbf{p}^{\prime}=-\mathbf{q}$. Thus,
	\begin{eqnarray}
		\Gamma^{}(f_{B} \rightarrow f_A \gamma)=\frac{1}{32\pi^2 m_B}\int\frac{d^{3}\mathbf{q}}{\bar{E}_{\mathbf{p}^{\prime}}E_{\mathbf{q}}} \delta(m_B-\bar{E}_{\mathbf{p}^{\prime}}-E_{\mathbf{q}})\lvert\overline{\mathcal{M}}(f_B\rightarrow f_A\gamma)\lvert^2 \Big\lvert_{\mathbf{p}^{\prime}\rightarrow-\mathbf{q}},
	\end{eqnarray}
being $\bar{E}_{\mathbf{p}^{\prime}}=\sqrt{m_A^{2}+\lvert \mathbf{q}\lvert^{2}}$. Additionally, the $\lvert\overline{\mathcal{M}}(f_B\rightarrow f_A\gamma)\lvert^2$ quantity  can be expressed in terms of masses $m_{A}$, $m_{B}$, and dot (cross) products between the 3-vector $\mathbf{q}$ and the real or imaginary parts of $\mathbf{e}^{AB}_{f}$ and $\mathbf{b}^{AB}_{f}$. Notice that these vector products depend on the components of $\mathbf{q}$, which are integration variables. In this regard, the phase-space integration may become cumbersome due to the choice of the reference frame in which the volume element is integrated. For simplicity, we work the scenario where $\mathbf{e}^{AB}_{f}$ and $\mathbf{b}^{AB}_{f}$ are orthogonal in the rest frame of the decaying particle $f_B$. Moreover, since the phase-space volume is Lorentz invariant, we adopt the reference frame defined by the triad conformed by the unitary vectors along the $\mathbf{e}^{AB}_{f}$, $\mathbf{b}^{AB}_{f}$, and $\mathbf{e}^{AB}_{f}\times\mathbf{b}^{AB}_{f}$ vectors; this reference frame is illustrated in Fig. \ref{SR2B}.
	On this basis, it follows that the decay rate is now
	\begin{eqnarray}
		\Gamma^{}(f_{B} \rightarrow f_A \gamma)=\frac{1}{64\pi^2}\frac{m_B^2-m_A^2}{m_B^{3}}\int_{0}^{2\pi}d\phi_\gamma\int_{0}^{\pi}d\theta_\gamma\sin\theta_\gamma \, \lvert\overline{\mathcal{M}}(f_B\rightarrow f_A\gamma)\lvert^2.
	\end{eqnarray}
	Since the integration extends over the full angular distribution, the decay width $\Gamma^{}(f_{B} \rightarrow f_A \gamma)$ will depend exclusively on the particle masses and the magnitudes of both real or imaginary parts of $\mathbf{e}^{AB}_{f}$ and $\mathbf{b}^{AB}_{f}$. The corresponding branching ratio is established as follows
	\begin{eqnarray}
		\mathrm{BR}^{YSME}(f_{B} \rightarrow f_A \gamma)=\frac{\Gamma(f_{B} \rightarrow f_A \gamma)}{\Gamma_B},
	\end{eqnarray}
	where $\Gamma_B$ is the total decay width of the $f_B$ fermion. After several manipulations of tensor algebra, it is found that, for the cases of pure real parameters ($\mathcal{A}^{AB}_f=0$) or pure imaginary parameters ($\mathcal{V}^{AB}_f=0$), the respective BRs can be expressed as:
	\begin{eqnarray}
		\mathrm{BR}^{YSME}_{\mathrm{Re}}(f_{B} \rightarrow f_A \gamma)&=& c_{R_1}^{AB}\left\lvert \mathrm{Re}\left(\mathbf{b}^{AB}_{f}\right) \right\lvert^{2}+c_{R_2}^{AB}\left\lvert \mathrm{Re}\left(\mathbf{e}^{AB}_{f}\right) \right\lvert^{2},\\
		\mathrm{BR}^{YSME}_{\mathrm{Im}}(f_{B} \rightarrow f_A \gamma)&=& c_{I_1}^{AB}\left\lvert \mathrm{Im}\left(\mathbf{b}^{AB}_{f}\right) \right\lvert^{2}+c_{I_2}^{AB}\left\lvert \mathrm{Im}\left(\mathbf{e}^{AB}_{f}\right) \right\lvert^{2},
	\end{eqnarray}
	where the coefficients $c_{I_{1,2}}^{AB}$, $c_{R_{1,2}}^{AB}$ are real and their explicit expressions are given by
	\begin{eqnarray}
		c_{R_{1}}^{AB}=&c_{I_{1}}^{AB}=&\frac{1}{411}\frac{v^2}{\Gamma_{B}m_{B}}\left(3-2\frac{m_A^2}{m_B^2}-\frac{m_A^4}{m_B^4}\right),\\
		c_{R_{2}}^{AB}=&c_{I_{2}}^{AB}=&\frac{1}{411}\frac{v^2}{\Gamma_{B}m_{B}}\left(5-4\frac{m_A^2}{m_B^2}-\frac{m_A^4}{m_B^4}\right),
	\end{eqnarray}
and
	\begin{eqnarray}
		\left\lvert \mathrm{Re}\left(\mathbf{c}^{AB}_{f}\right) \right\lvert^{2}&=&\mathrm{Re}\left(\mathbf{c}^{AB}_{f}\right)\cdot \mathrm{Re}\left(\mathbf{c}^{AB}_{f}\right),\\
		\left\lvert \mathrm{Im}\left(\mathbf{c}^{AB}_{f}\right) \right\lvert^{2}&=&\mathrm{Im}\left(\mathbf{c}^{AB}_{f}\right)\cdot \mathrm{Im}\left(\mathbf{c}^{AB}_{f}\right),
	\end{eqnarray}
	with $\textbf{c}=\textbf{e},\textbf{b}$. The numerical values for the $c_{R_{1,2}}^{AB}$ coefficients are given in Table \ref{CABCoef2Bodies}. The muon and tau decay widths are derived from their lifetimes, the corresponding masses and lifetimes were taken from Ref.~\cite{PDG}.

	\begin{table}[th!]
		\caption{Numerical values of the $	c_{R_{1,2}}^{AB}$ coefficients.}\label{CABCoef2Bodies}
		\centering
		\begin{tabular}{c|c|c}
			\hline\hline
			$\tau^{-}\rightarrow\mu^{-}\gamma$&$c_{R_{1}}^{\mu\tau}=1.10\times 10^{14}$&$c_{R_{2}}^{\mu\tau}=1.83\times 10^{14}$\\
			$\tau^{-}\rightarrow e^{-}\gamma$&$c_{R_{1}}^{e\tau}=1.10\times 10^{14}$&$c_{R_{2}}^{e\tau}=1.84\times 10^{14}$\\
			$\mu^{-}\rightarrow e^{-}\gamma$&$c_{R_{1}}^{e\mu}=1.40\times 10^{22}$&$c_{R_{2}}^{e\mu}=2.33\times 10^{22}$ \\
			\hline\hline
		\end{tabular}
	\end{table}
	
\begin{figure}[ht!]
		\centering
		
		\subfloat[]{%
			\includegraphics[width=0.41\textwidth]{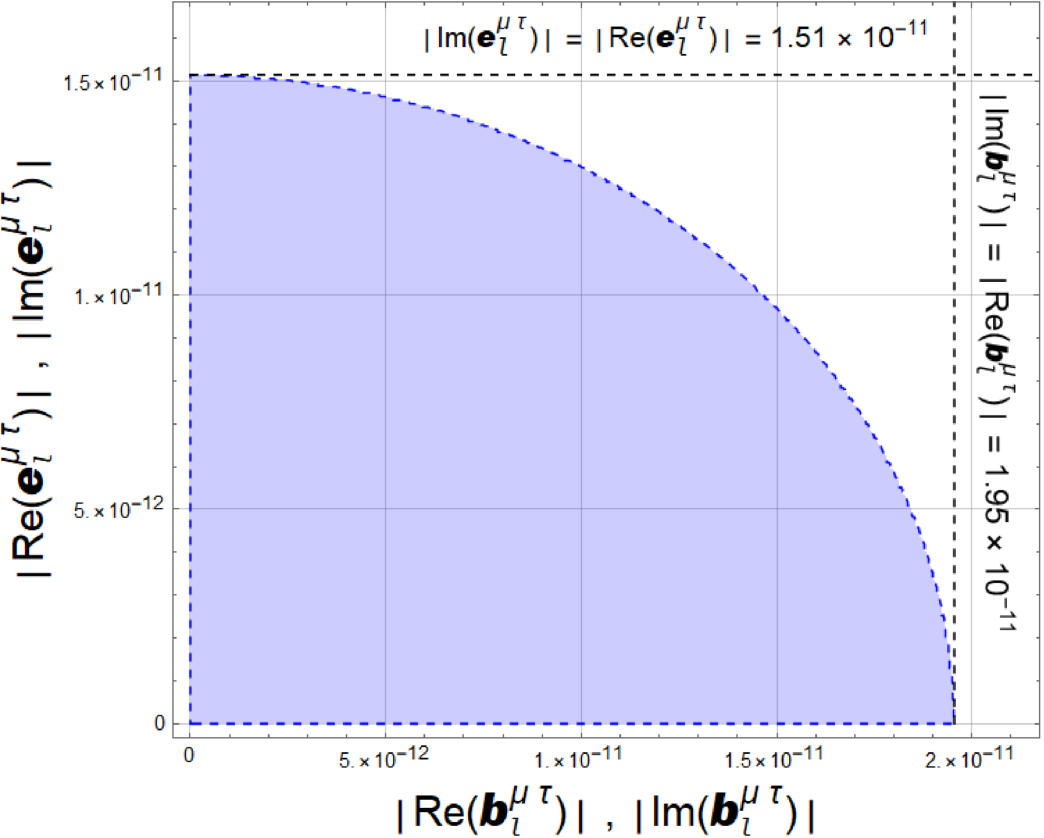}
		}
		\hspace{0.01\textwidth}
		\subfloat[]{%
			\includegraphics[width=0.41\textwidth]{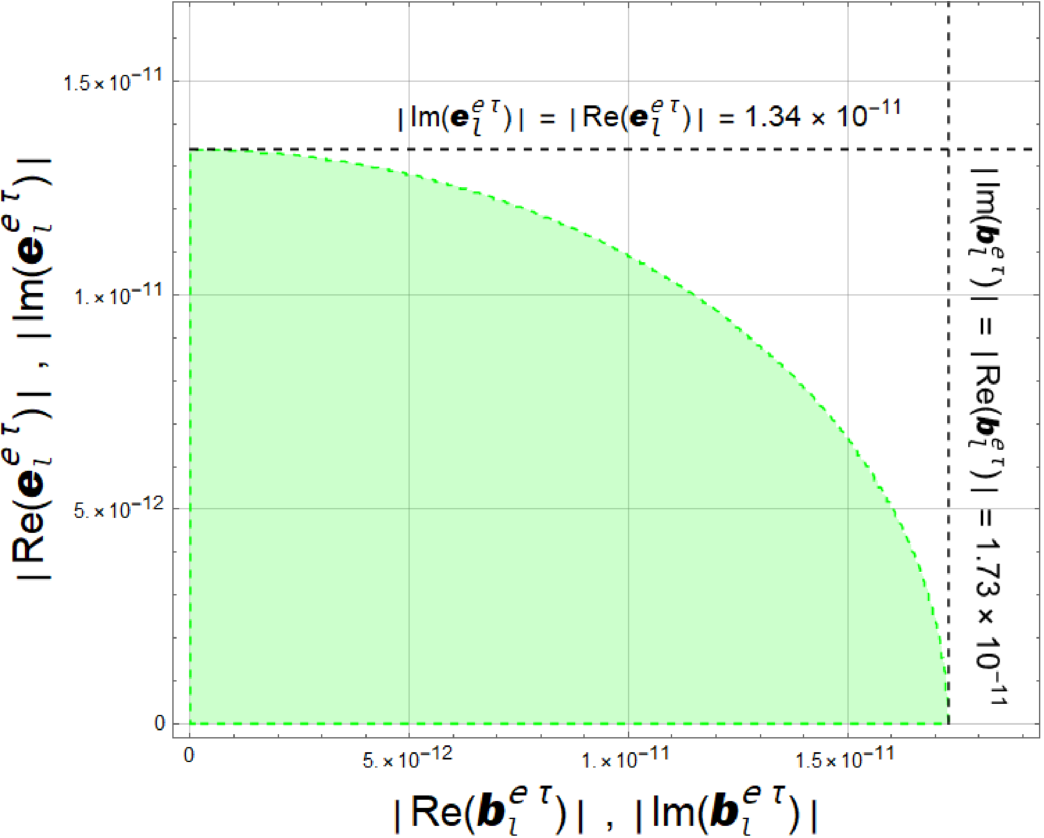}
		}
		\hspace{0.01\textwidth}
		\subfloat[]{%
			\includegraphics[width=0.41\textwidth]{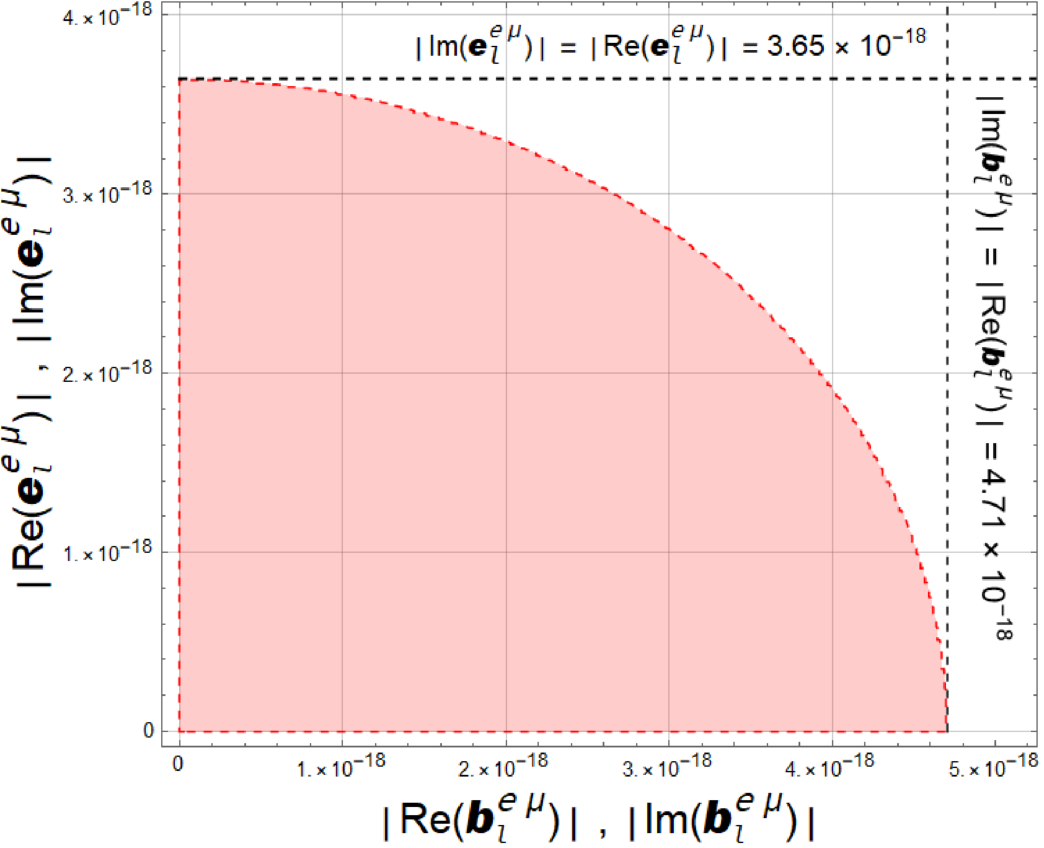}
		}
		
		\caption{Allowed  regions for $\lvert\mathbf{b}^{AB}_{l}\lvert$, $\lvert\mathbf{e}^{AB}_{l}\lvert$ stem from current experimental bounds for: (a) $\tau\rightarrow\mu\gamma$, (b) $\tau\rightarrow e \gamma$, and (c) $\mu\rightarrow e\gamma$.}
		\label{2BodyRegionsVSL}
	\end{figure}

\noindent In order to estimate LV effects, it is proposed that any new physics effect must be smaller than current experimental bounds on LFV processes as seen through their branching ratios. Thus,
	\begin{eqnarray}\label{ebCondition2b}
		\mathrm{BR}^{YSME}_{\mathrm{Re},\mathrm{Im}}(f_{B} \rightarrow f_A \gamma)\leq \mathrm{BR}^{Exp}(f_{B} \rightarrow f_A \gamma)\, ,
	\end{eqnarray}
where $\mathrm{BR}^{Exp}(f_{B} \rightarrow f_A \gamma)$ is expressly stated in the first column of Table \ref{ExpBounds}, accordingly for each decay mode. Thereafter, Eq.~(\ref{ebCondition2b}) restricts the maxima of the values allowed for $\lvert\mathbf{b}^{AB}_{f}\lvert$ and $\lvert\mathbf{e}^{AB}_{f}\lvert$. The pattern imposed by the proposed method for bounding the LV effects can be visualized in a two-dimensional plot ($\lvert\mathbf{b}^{AB}_{f}\lvert$ versus $\lvert\mathbf{e}^{AB}_{f}\lvert$), just as it is shown in Fig. \ref{2BodyRegionsVSL}. It should be noted that the resulting region delimited by the ellipse considers as variables $\lvert\mathrm{Re}(\mathbf{b}^{AB}_{f})\lvert$ ($\lvert\mathrm{Im}(\mathbf{b}^{AB}_{f})\lvert$) versus 	$\lvert\mathrm{Re}(\mathbf{e}^{AB}_{f})\lvert$ ($\lvert\mathrm{Im}(\mathbf{e}^{AB}_{f})\lvert$).
	
The maxima of the permitted values for  $\lvert \mathbf{b}^{AB}_{l}\lvert$, $\lvert \mathbf{e}^{AB}_{l}\lvert$ in the CM frame of the decaying particle are presented in Table \ref{MaxEB2bodies}. For the cases where the $(Y_l)^{AB}_{\mu\nu}$ parameters are either real or pure imaginary, we obtain constraints of similar order of magnitude in both scenarios. In the literature, it is reported a constraint for the LV parameter $\lvert\lvert\mathrm{Re}(\mathbf{e}^{\mu\tau}_{l})\lvert^2-\lvert\mathrm{Re}(\mathbf{b}^{\mu\tau}_{l})\lvert^2\lvert$ \cite{AhuatziAvendano2020ppm}, being explicitly $\lvert\lvert\mathrm{Re}(\mathbf{e}^{\mu\tau}_{l})\lvert^2-\lvert\mathrm{Re}(\mathbf{b}^{\mu\tau}_{l})\lvert^2\lvert\sim 10^{-15}$, whereas our estimation is of the order of $10^{-22}$, which is seven orders of magnitude more restrictive. Moreover, in contrast to Ref. \cite{AhuatziAvendano2020ppm}, we explicitly consider Lorentz contractions with the momenta of the particles involved, which were neglected in that work, where the resulting contributions of this contractions stand out around $60\%$ on all the partial decay widths.
	\begin{table}[th!]
		\caption{Upper bounds on $\lvert \mathbf{b}^{AB}_{l}\lvert$ and $\lvert \mathbf{e}^{AB}_{l}\lvert$ derived from the two-body $f_B\rightarrow f_A\gamma$ decay .}\label{MaxEB2bodies}
		\centering
		\begin{tabular}{ccc}
			\hline\hline
			$\lvert \mathbf{b}^{\mu\tau}_{l}\lvert\leq 1.95\times 10^{-11}$ 	&& $\lvert\mathbf{e}^{\mu\tau}_{l}\lvert\leq 1.51\times 10^{-11}$ \\
			$\lvert \mathbf{b}^{e\tau}_{l}\lvert\leq 1.73\times 10^{-11}$  && $\lvert\mathbf{e}^{e\tau}_{l}\lvert\leq 1.34\times 10^{-11}$ \\
			$\lvert \mathbf{b}^{e\mu}_{l}\lvert\leq 4.71\times 10^{-18}$ 	&& $\lvert\mathbf{e}^{e\mu}_{l}\lvert\leq 3.65\times 10^{-18}$\\
			\hline\hline
		\end{tabular}
	\end{table}
	
	\pagebreak
	
	\subsection{The three-body $f_B \rightarrow f_{A} f_C \bar{f}_C$ decay}
\begin{figure}[ht!]
		\includegraphics[trim= -30mm 230mm 0mm 25mm, scale=0.6,clip]{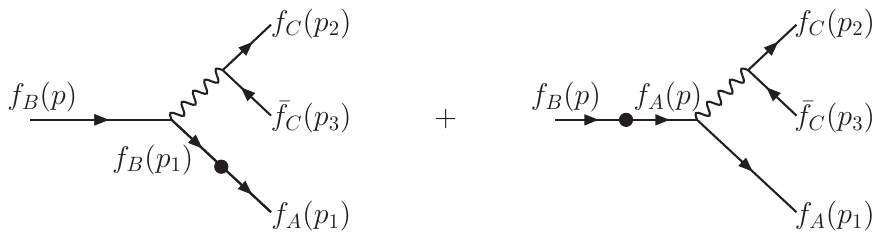}
		\caption{First-order LV corrections to the $f_B \rightarrow f_{A} f_C \bar{f}_C$ decay.}\label{BToACC}
	\end{figure}

The first-order contributions of the MEY sector to the $f_{B} \rightarrow f_{A} f_C \bar{f}_C$ process are calculated via the Feynman diagrams presented in Fig.~\ref{BToACC}.
	
\begin{figure}[ht!]
	\centering
	
	\subfloat[]{%
		\includegraphics[width=0.41\textwidth]{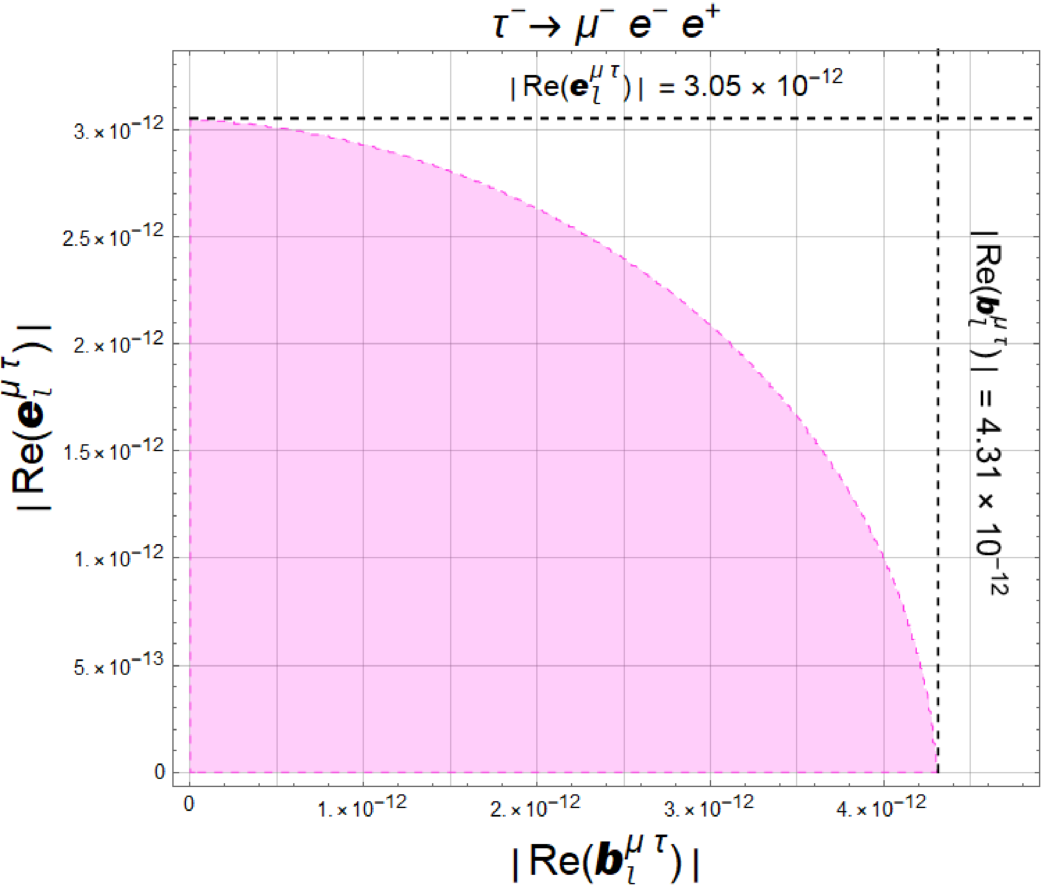}
	}
	\hspace{0.01\textwidth}
	\subfloat[]{%
		\includegraphics[width=0.41\textwidth]{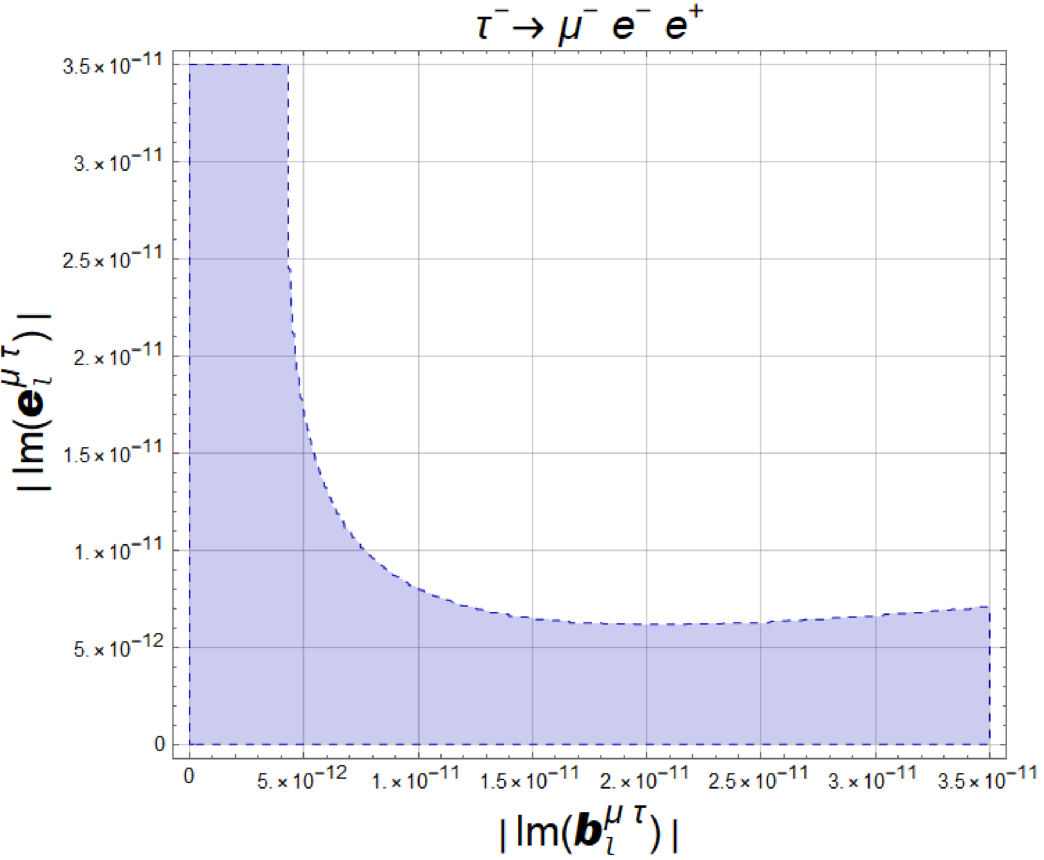}
	}
	\hspace{0.01\textwidth}
	\caption{Allowed  regions for  (a) $\lvert\mathrm{Re}(\mathbf{b}^{\mu\tau}_{l})\lvert$ vs. $\lvert\mathrm{Re}(\mathbf{e}^{\mu\tau}_{l})\lvert$ and (b) $\lvert\mathrm{Im}(\mathbf{b}^{\mu\tau}_{l})\lvert$ vs. $\lvert\mathrm{Im}(\mathbf{e}^{\mu\tau}_{l})\lvert$, derived from the experimental bound for the $\tau^{-}\rightarrow \mu^{-} e^{-} e^{+}$ decay.}
	\label{3BodyRegionsVSL}
\end{figure}

	\noindent The associated amplitude can be written as
	\begin{eqnarray}\label{GBToACC}
		\mathcal{M}(f_B\rightarrow f_A f_C \bar{f}_C)= \frac{2 i\, v \, \pi \alpha \, Q^2_{f}}{(p_2+p_3)^{2}}\,\,  \bar{u}^{s^{\prime}}(p_1,m_A)\left[ T^{AB}_{f}\left(\frac{\slashed{p}_1+m_B}{p_{1}^{2}-m_B^2}\right)\gamma_{\mu}+\gamma_{\mu}\left(\frac{\slashed{p}+m_A}{p^{2}-m_A^2}\right)T^{AB}_{f}\right]u^{s}(p,m_B)\nonumber\\
		\times \bar{u}^{r^{\prime}}(p_2,m_C)\gamma^{\mu}v^{r}(p_3,m_C).\qquad\qquad\qquad\qquad\qquad\qquad\qquad\qquad
	\end{eqnarray}
The respective BR is given by
	\begin{eqnarray}
		\mathrm{Br}(f_B\rightarrow f_A f_C \bar{f}_C)=\frac{1}{2m_B}\frac{1}{\Gamma_B}\int\left(\frac{1}{2E_{\mathbf{p}_1}}\frac{d^{3}\mathbf{p}_1}{(2\pi)^3}\right)\left(\frac{1}{2E_{\mathbf{p}_2}}\frac{d^{3}\mathbf{p}_2}{(2\pi)^3}\right)\left(\frac{1}{2E_{\mathbf{p}_3}}\frac{d^{3}\mathbf{p}_3}{(2\pi)^3}\right)\nonumber\\
		\times (2\pi)^4 \delta^{(4)}(p-p_1-p_2-p_3) \lvert\overline{\mathcal{M}}(f_B\rightarrow f_A f_C \bar{f}_C)\lvert^2.
	\end{eqnarray}
Just as in the case of the two-body decay, the volume element $d^{3}\mathbf{p_3}$ is integrated at the rest frame of the decaying particle $f_B$, whereas the phase-space volume elements $d^{3}\mathbf{p_2}$ and $d^{3}\mathbf{p_1}$ are integrated in the reference frame depicted in Fig.~\ref{SR2B}. Therefore, by using the same scenario for the LV  parameters as the one introduced for the case of two-body decay, the BR can be expressed as
	\begin{eqnarray}
		\mathrm{BR}^{YSME}_{\mathrm{Re}}(f_B\rightarrow f_A f_C \bar{f}_C)= d_{R_1}^{AB}\left\lvert \mathrm{Re}\left(\mathbf{b}^{AB}_{f}\right) \right\lvert^{2}+d_{R_2}^{AB}\left\lvert \mathrm{Re}\left(\mathbf{b}^{AB}_{f}\right) \right\lvert\left\lvert \mathrm{Re}\left(\mathbf{e}^{AB}_{f}\right) \right\lvert+d_{R_3}^{AB}\left\lvert \mathrm{Re}\left(\mathbf{e}^{AB}_{f}\right) \right\lvert^{2},\\
		\mathrm{BR}^{YSME}_{\mathrm{Im}}(f_B\rightarrow f_A f_C \bar{f}_C)= d_{I_1}^{AB}\left\lvert \mathrm{Im}\left(\mathbf{b}^{AB}_{f}\right) \right\lvert^{2}+d_{I_2}^{AB}\left\lvert \mathrm{Im}\left(\mathbf{b}^{AB}_{f}\right) \right\lvert\left\lvert \mathrm{Im}\left(\mathbf{e}^{AB}_{f}\right) \right\lvert+d_{I_3}^{AB}\left\lvert \mathrm{Im}\left(\mathbf{e}^{AB}_{f}\right) \right\lvert^{2}.
	\end{eqnarray}
The explicit analytical expressions for the $d_{R_i}^{AB}$ and $d_{I_i}^{AB}$ coefficients are extremely lengthy and do not provide additional physical insight, however, it can be determined by using Eqs.~(\ref{A2})-(\ref{A10}). In specific, for all the processes in question, only the $\tau^{-}\rightarrow \mu^{-} e^{-} e^{+}$ decay provides a stringent constraint for the LV parameters.  Consequently, the $\mathrm{BR}(\tau^{-}\rightarrow \mu^{-} e^{-} e^{+})$ is now
	\begin{eqnarray}
		\mathrm{BR}^{YSME}_{\mathrm{Re}}(\tau^{-}\rightarrow \mu^{-} e^{-} e^{+})= \left(+9.70\left\lvert \mathrm{Re}\left(\mathbf{b}^{\mu\tau}_{l}\right) \right\lvert^{2}-1.28\left\lvert \mathrm{Re}\left(\mathbf{b}^{\mu\tau}_{l}\right) \right\lvert\left\lvert \mathrm{Re}\left(\mathbf{e}^{\mu\tau}_{l}\right) \right\lvert+19.4\left\lvert \mathrm{Re}\left(\mathbf{e}^{\mu\tau}_{l}\right) \right\lvert^{2}\right)\times 10^{14},\\
		\mathrm{BR}^{YSME}_{\mathrm{Im}}(\tau^{-}\rightarrow \mu^{-} e^{-} e^{+})=\left(-4.46\left\lvert \mathrm{Im}\left(\mathbf{b}^{\mu\tau}_{l}\right) \right\lvert^{2}+29.6\left\lvert \mathrm{Im}\left(\mathbf{b}^{\mu\tau}_{l}\right) \right\lvert\left\lvert \mathrm{Im}\left(\mathbf{e}^{\mu\tau}_{l}\right) \right\lvert-2.17\left\lvert \mathrm{Im}\left(\mathbf{e}^{\mu\tau}_{l}\right) \right\lvert^{2}\right)\times 10^{13} .
	\end{eqnarray}
By comparing with the respective experimental BR, in the same way considered for the procedure proposed in Eq.~(\ref{ebCondition2b}), it follows that
	\begin{eqnarray}
		\mathrm{BR}^{YSME}_{\mathrm{Re},\mathrm{Im}}(\tau^{-}\rightarrow \mu^{-} e^{-} e^{+})\leq 1.8\times 10^{-8}\, ,
	\end{eqnarray}
which provides regions delimited by an ellipse or a hyperbole in plots ($\lvert\mathbf{b}^{AB}_{f}\lvert$ versus $\lvert\mathbf{e}^{AB}_{f}\lvert$) (see Fig.~\ref{3BodyRegionsVSL}).

\noindent Thus, for pure-real LV parameters, we obtain upper bounds for the $(Y_{l})^{\mu\tau}$ components
\begin{eqnarray}\label{MaxEB3bodies}
	\lvert \mathbf{b}^{\mu\tau}_{l}\lvert\leq 4.31\times 10^{-12}, \qquad \lvert \mathbf{e}^{\mu\tau}_{l}\lvert\leq 3.05\times 10^{-12}.
\end{eqnarray}
\begin{figure}
	\centering
	\includegraphics[scale=0.41]{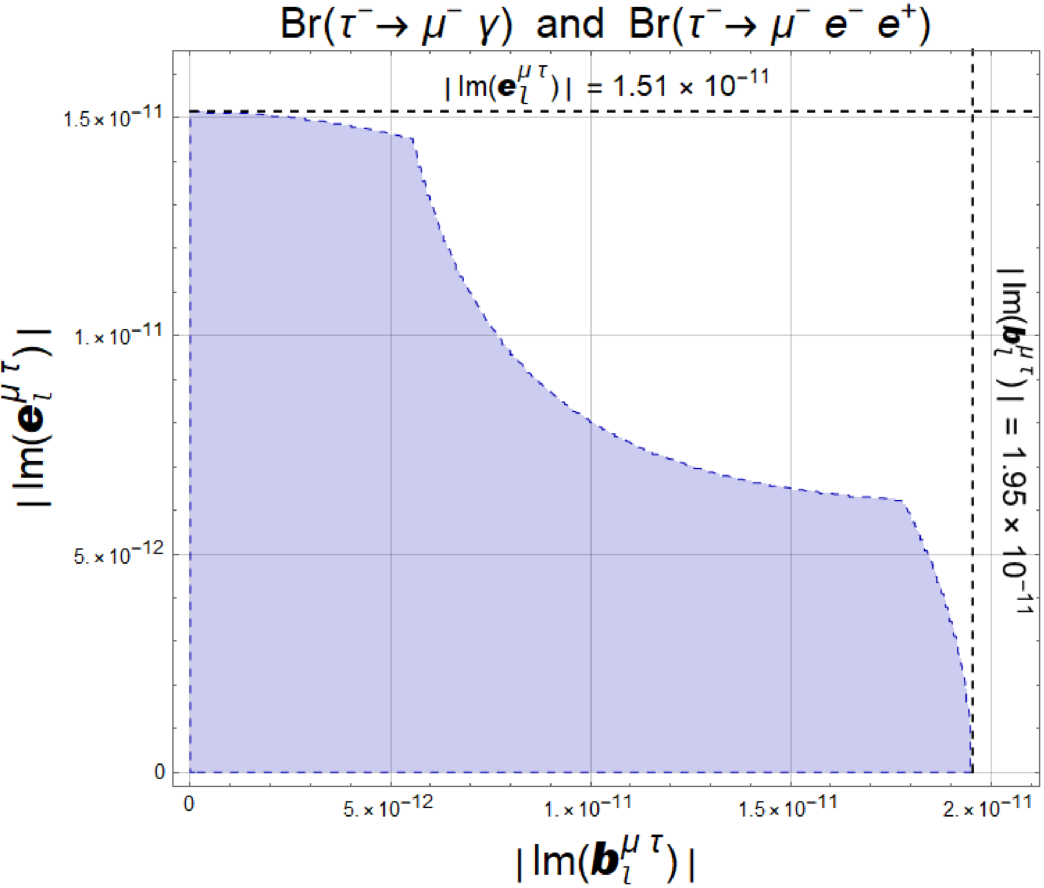}
	\caption{Allowed region for the $\lvert\mathrm{Im}(\mathbf{b}^{\mu\tau}_{l})\lvert$ and $\lvert\mathrm{Im}(\mathbf{e}^{\mu\tau}_{l})\lvert$ parameters derived form the combined results obtained with the $\tau\rightarrow\mu\gamma$ and $\tau^{-}\rightarrow\mu^{-}e^{+}e^{-}$ decay processes.}\label{3and2BodyRegionsVSL}
\end{figure}
For pure-imaginary LV parameters, an unbounded region is obtained (see  Fig.~\ref{3BodyRegionsVSL} (b)) and no direct constraints can be deduced from this scenario. Nevertheless, by superposing the plots $\lvert\mathrm{Im}(\mathbf{b}^{\mu\tau}_{l})\lvert$ vs. $\lvert\mathrm{Im}(\mathbf{e}^{\mu\tau}_{l})\lvert$ for both two-body and three-body decays,  we can discard the unbounded region and to establish a bounded region for the LV parameters in question, as it is shown in Fig.~\ref{3and2BodyRegionsVSL}. Since there are no modifications to the upper limits in the resulting region, we will retain the upper bounds previously calculated from the two-body decay.

Finally, the bounds given in both Table \ref{MaxEB2bodies} and Eq.~(\ref{MaxEB3bodies}) can be translated into constrains on the $(Y_l)^{AB}_{\alpha\beta}$ components through the relations established in Eqs. (\ref{ebDef1}) and (\ref{ebDef2}). Thus, all the corresponding LV bounds in the rest frame of the decaying particles are presented in Tab. \ref{MaxYRestFrame}.
\begin{table}[th!]
		\caption{Upper bounds for the $(Y_l)^{AB}_{0i}$ and $(Y_l)^{AB}_{ij}$ components; these bounds were computed  in the rest frame of the decaying particles.}\label{MaxYRestFrame}
	\begin{tabular}{c|cc|cc|c}
		\hline \hline
		\multicolumn{2}{c}{Pure real} &  &  & \multicolumn{2}{c}{Pure imaginary} \\
		\hline
		$(Y_l)^{\mu\tau}_{0i}(Y_l)^{\mu\tau}_{0i}<9.30\times 10^{-24}$             &  $(Y_l)^{\mu\tau}_{ij}(Y_l)^{\mu\tau}_{ij}<3.72\times 10^{-23}$           &  &  &$(Y_l)^{\mu\tau}_{0i}(Y_l)_{0i}^{*\mu\tau}<2.28\times 10^{-22}$&$(Y_l)^{\mu\tau}_{ij}(Y_l)_{ij}^{*\mu\tau}<7.60\times 10^{-22}$                 \\
		$(Y_l)^{e\tau}_{0i}(Y_l)^{e\tau}_{0i}<1.80\times 10^{-22}$             &  $(Y_l)^{e\tau}_{ij}(Y_l)^{e\tau}_{ij}<5.99\times 10^{-22}$           &  &  &$(Y_l)^{e\tau}_{0i}(Y_l)_{0i}^{*e\tau}<1.80\times 10^{-22}$&$(Y_l)^{e\tau}_{ij}(Y_l)_{ij}^{*e\tau}<5.99\times 10^{-22}$                 \\
		$(Y_l)^{e\mu}_{0i}(Y_l)^{e\mu}_{0i}<1.33\times 10^{-35}$             &  $(Y_l)^{e\mu}_{ij}(Y_l)^{e\mu}_{ij}<4.44\times 10^{-35}$           &  &  &$(Y_l)^{e\mu}_{0i}(Y_l)_{0i}^{*e\mu}<1.33\times 10^{-35}$&$(Y_l)^{e\mu}_{ij}(Y_l)_{ij}^{*e\mu}<4.44\times 10^{-35}$                 \\
		\hline \hline
	\end{tabular}
\end{table}

\begin{table}[ht!]
	\caption{Upper bounds for the Lorentz-invariant scalars $(Y_l)^{AB}_{\alpha\beta}(Y_l)^{*BA\alpha\beta}$.}\label{MaxY}
\begin{tabular}{cc|cc}
	\hline\hline
	Pure real &  &  & Pure imaginary \\
	\hline
	$|(Y_l)^{ \mu\tau}_{\alpha\beta}(Y_l)^{*\tau\mu\alpha\beta}|<5.58\times 10^{-23}$       &  &  & 	$|(Y_l)^{ \mu\tau}_{\alpha\beta}(Y_l)^{*\tau\mu\alpha\beta}|<1.22\times 10^{-21}$               \\
	$|(Y_l)^{ e\tau}_{\alpha\beta}(Y_l)^{*\tau e\alpha\beta}|<9.58\times 10^{-22}$&  &  &  $|(Y_l)^{ e\tau}_{\alpha\beta}(Y_l)^{*\tau e\alpha\beta}|<9.58\times 10^{-22}$              \\
	$|(Y_l)^{ e\mu}_{\alpha\beta}(Y_l)^{*\mu e\alpha\beta}|<7.10\times 10^{-35}$&  &  &  $|(Y_l)^{ e\mu}_{\alpha\beta}(Y_l)^{*\mu e\alpha\beta}|<7.10\times 10^{-35}$              \\
	\hline\hline
\end{tabular}
\end{table}

\noindent Since scalars of the form $(Y_f)^{AB}_{\alpha\beta}(Y_f)^{*BA\, \alpha\beta}$ are Lorentz invariants in the flavor space (see Eq.~(\ref{A11})), we can use the bounds above obtained to establish limits on their values for any reference frame, which can be appreciated in Table~\ref{MaxY}.

\subsection{Second-order contributions}
\begin{figure}[th!]
	\includegraphics[trim= 0mm 235mm 0mm 25mm, scale=0.75,clip]{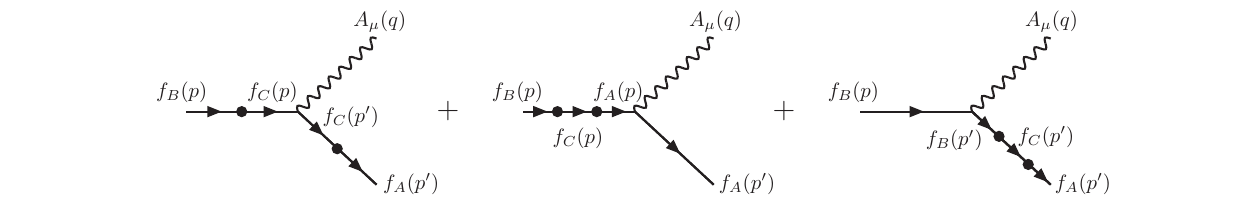}
	\caption{Second-order corrections to the LV parameters $Y_{f}^{AB}$ on the $f_{B}f_{A}\gamma$ vertex.}\label{BToAGO2}
\end{figure}
In order to search for dipolar-transition form factors, we compute second-order insertions in the $f_Bf_A\gamma$ coupling at tree level, which is represented by the sum of the Feynman diagrams shown in Fig. \ref{BToAGO2}. In this manner, the vertex function results in
\begin{eqnarray}
	\Gamma^{O(2)f_Bf_A\gamma}_{\mu}=\displaystyle i e Q \bar{u}^{s^{\prime}}(p^\prime,m_A)\Gamma^{O(2)AB}_{\mu}u^{s}(p,m_B),
\end{eqnarray}
where
\begin{eqnarray}\label{SeconOrderVertex}
	\Gamma^{O(2)AB}_{\mu}&=& (T^{AB}_{V})_{\mu} -(T^{AB}_{A})_{\mu}\gamma_5+\gamma_{\mu}(g^{AB}_{V}-g^{AB}_{A}\gamma_5)+i\sigma_{\mu\lambda}q_{\rho}\left[\frac{(F^{AB}_{M})^{\lambda\rho}}{m_A+m_B}-\frac{(F_{E}^{AB})^{\lambda\rho}}{m_B-m_A}\gamma_5\right]\nonumber\\
	&&-i \epsilon_{\mu\lambda\rho\tau}\gamma^{\lambda}\Big[(R^{AB}_V)^{\rho\tau}-(R^{AB}_A)^{\rho\tau}\gamma_5\Big]+\gamma^{\lambda}\Big[(S^{AB}_V)_{\mu\lambda}-(S^{AB}_A)_{\mu\lambda}\gamma_5 \Big]\nonumber\\
	 &&+i\sigma_{\lambda\rho}\left[\frac{(\hat{F}^{AB}_{M})^{\mu\lambda\rho}}{m_A+m_B}-\frac{(\hat{F}_{E}^{AB})^{\mu\lambda\rho}}{m_B-m_A}\gamma_5\right].
\end{eqnarray}
In this case, $q=p^\prime-p$. The explicit form for all the Lorentz-tensor structures present in Eq.~(\ref{SeconOrderVertex}) are given in Appendix \ref{AppendixLorentzStructures}.
This amplitude is gauge invariant since it satisfies the Ward identity $q^{\mu}\Gamma^{O(2)A\, B}_{\mu}=0$.

Although in Eq.~(\ref{SeconOrderVertex}) there is no presence of dipolar-transition structures ($\sigma^{\mu\nu}q_\nu$), it is of worth to appreciate generalized vector and axial flavor-changing neutral currents, where their respective dimensionless  couplings $g_{V}^{AB}$ and $g_{A}^{AB}$ are
\begin{eqnarray}
	g_{V}^{AB}&=&\frac{2v^2}{ m_{B A}^2 m_{C B}^2 m_{A C}^2}\Big[ m_{B A}^2  \left(\mathcal{V}^{AC}_f\right)_{\lambda \, q}\left(\mathcal{V}^{CB}_f\right)^{\lambda \, q}+ m_{A C}^2 \left(\mathcal{V}^{AC}_f\right)_{\lambda \, q}\left(\mathcal{V}^{CB}_f\right)^{\lambda \, P} \nonumber\\
	&&\qquad\qquad\qquad\qquad\qquad+  m_{C B}^2 \left(\mathcal{V}^{AC}_f\right)_{\lambda \, P}\left(\mathcal{V}^{CB}_f\right)^{\lambda \, q}-(\mathcal{V}\rightarrow \mathcal{A})\Bigg],
\end{eqnarray}
\begin{eqnarray}
	g_{A}^{AB}&=&\frac{2v^2}{m_{B A}^2 m_{C B}^2 m_{A C}^2}\Big[ m_{B A}^2  \left(\mathcal{A}^{AC}_f\right)_{\lambda \, q}\left(\mathcal{V}^{CB}_f\right)^{\lambda \, q}+ m_{A C}^2 \left(\mathcal{A}^{AC}_f\right)_{\lambda \, q}\left(\mathcal{V}^{CB}_f\right)^{\lambda \, P} \nonumber\\
	&&\qquad\qquad\qquad\qquad\qquad+  m_{C B}^2 \left(\mathcal{A}^{AC}_f\right)_{\lambda \, P}\left(\mathcal{V}^{CB}_f\right)^{\lambda \, q}-(\mathcal{V}\leftrightarrow \mathcal{A})\Bigg],
\end{eqnarray}
with $m_{XY}^{2}=m_X^2-m_Y^2$, for $X,Y=A,B,C$. An interesting property is that $g_{V}^{AB}$ and $g_{A}^{AB}$ are invariant under observer Lorentz transformations, but not under particle transformations.

\section{Conclusions}\label{SecIV}
In this work, we calculated BRs for the LFV processes $f_B\rightarrow f_A\gamma$ and $f_B\rightarrow f_A f_C\bar{f}_C$ mediated by the Yukawa sector of the SME through the LV tensor $(Y_f)^{AB}_{\mu\nu}$. By comparing with the respective experimental bounds, parameter spaces for the LV couplings, $\lvert\mathbf{e}^{AB}_{f}\lvert$ and $\lvert\mathbf{b}^{AB}_{f}\lvert$, were determined. In order to perform the phase space integration for the different decays, some assumptions over the spatial orientation for the 3-vectors $\mathbf{e}^{AB}_{f}$ and $\mathbf{b}^{AB}_{f}$ associated to $(Y_f)^{AB}_{0i}$ and $(Y_f)^{AB}_{ij}$, respectively, were considered. We took $\mathbf{e}^{AB}_{f}$ and $\mathbf{b}^{AB}_{f}$ as either pure real or pure imaginary, establishing them mutually orthogonal in the rest frame of the decaying particle. It was found that $\lvert\mathbf{e}^{e\tau}_{l}\lvert$ or  $\lvert\mathbf{b}^{e\tau}_{l}\lvert$  are below of $10^{-11}$, whilst  $\lvert \mathbf{e}^{\mu\tau}_{l}\lvert$ or $\lvert \mathbf{b}^{\mu\tau}_{l}\lvert$ are below of  $10^{-12}$. The most stringent bounds corresponded to $\lvert \mathbf{e}^{e\mu}_{l}\lvert$ or $\lvert \mathbf{b}^{e\mu}_{l}\lvert$, being of the order of $10^{-18}$. It should be recalled that this work introduces a methodology that addresses new energy-momentum dependent terms coupled to background LV-tensor fields that were ignored in previous researches. Our upper bounds result more restrictive than one of those previously reported in the literature. In addition, it is presented new constraints for Lorentz-invariant scalars $(Y_l)^{AB}_{\alpha\beta}(Y_l)^{*BA\alpha\beta}$ in the flavor space. On the other hand, we computed second-order LV corrections to the $f_B f_A \gamma$ vertex, finding that these only induce generalized vector and axial transition currents and a variety of new Lorentz structures, where there is no presence of dipolar-transition structures.

For future investigations, the upper bounds for the different LV parameters that we estimated could be implemented in the study of LFV scattering processes, in which the spatial orientation of the momentum of the initial particles with respect to the background fields $\mathbf{e}^{AB}_{f}$ and $\mathbf{b}^{AB}_{f}$ could be manifest through their cross sections.

\acknowledgments
This work has been partially supported by SNII-Secihti and CIC-UMSNH. JMD thanks to the Secihti program Investigadoras e Investigadores por M\'exico, project 7009. OVH thanks to Secihti for Postdoctoral support.

\appendix

\section{Lorentz contractions}\label{AppendixA}
In this appendix, we present the scalar products that arise in this work. For convenience, we denote the LV tensors $(Y_f)^{AB}_{\mu\nu}$, $(Y_f)^{*BA}_{\mu\nu}$, $(\tilde{Y}_f)^{AB}_{\mu\nu}$, $(\tilde{Y}_f)^{*BA}_{\mu\nu}$ as $(\mathcal{X}^{AB})_{\mu\nu}$ or $(\mathcal{Z}^{AB})_{\mu\nu}$. As mentioned in Introduction, the six independent components of $(Y_f)^{AB}_{\mu\nu}$ can be expressed in terms of $(\mathbf{e}^{AB}_{f})^{i}$ and $(\mathbf{b}^{AB}_{f})^{i}$ (see Eq.~(\ref{ebDef1})). Thus, the $(\mathcal{X}^{AB})_{\mu\nu}$ tensor can be associated with two three-vectors $(\mathbf{X}^{AB})^{i}$ and $(\mathbf{x}^{AB})^{i}$ through
\begin{eqnarray}
	(\mathcal{X}^{AB})_{0 i} &=& (\mathbf{X}^{AB})_{i} , \qquad	(\mathcal{X}^{AB})_{ij}=\epsilon^{ijk}(\mathbf{x}^{AB})^{k}.
\end{eqnarray}
where $\mathbf{X}^{AB}$ and $\mathbf{x}^{AB}$ correspond to $\mathbf{e}^{AB}$ or $\mathbf{b}^{AB}$, the corresponding relations are summarized in Table \ref{XYvectors}, see also Eqs.~(\ref{ebDef1}) and (\ref{ebDef2}).
\begin{table}[th!]
		\caption{Associated  $(\mathbf{e}^{AB}_{f})^{i}$ or $(\mathbf{b}^{AB}_{f})^{i}$ components for each $(\mathcal{X}^{AB})_{\mu\nu}$ tensor.}\label{XYvectors}
	\centering
	\begin{tabular}{ccccc}
		$(\mathcal{X}^{AB})_{\mu\nu}$&& $(\mathbf{X}^{AB})^{i}$&&$(\mathbf{x}^{AB}_{f})^{i}$\\
		\hline\hline\\
		\vspace{0.1cm}$(Y_f)^{AB}_{\mu\nu}$&& $+(\mathbf{e}^{AB}_{f})^{i}$ && $+(\mathbf{b}^{AB}_{f})^{i}$\\
		\vspace{0.1cm}$(Y_f)^{*BA}_{\mu\nu}$&&$+(\mathbf{e}^{BA*}_{f})^{i}$ &&$+(\mathbf{b}^{BA*}_{f})^{i}$\\
		\vspace{0.1cm}$(\tilde{Y}_f)^{AB}_{\mu\nu}$&& $+(\mathbf{b}^{AB}_{f})^{i}$ && $-(\mathbf{e}^{AB}_{f})^{i}$\\
		\vspace{0.1cm}$(\tilde{Y}_f)^{*BA}_{\mu\nu}$&&$+(\mathbf{b}^{BA*}_{f})^{i}$ &&$-(\mathbf{e}^{BA*}_{f})^{i}$\\
		\hline\hline\\
	\end{tabular}
\end{table}
In the $\mathbf{X}^{AB}$ and $\mathbf{x}^{AB}$ basis, the Lorentz scalar products that appear in this paper are given by
%%%%%%%%%%%%%%%%%%%%%%%%%%%%%%%%%%%%%%%%%%
%%%%%%%%%%%%% Type 1 %%%%%%%%%%%%%%%%%%%%%
%%%%%%%%%%%%%%%%%%%%%%%%%%%%%%%%%%%%%%%%%%
\begin{eqnarray}\label{A2}
	(\mathcal{X}^{AB})_{\mu\nu}(\mathcal{Z}^{AB})^{\mu\nu}&=&-2 \, \mathbf{X}^{AB}\cdot \mathbf{Z}^{AB}+2\, \mathbf{x}^{AB}\cdot \mathbf{z}^{AB},
\end{eqnarray}
%%%%%%%%%%%%%%%%%%%%%%%%%%%%%%%%%%%%%%%%%%
%%%%%%%%%%%%% Type 2 %%%%%%%%%%%%%%%%%%%%%
%%%%%%%%%%%%%%%%%%%%%%%%%%%%%%%%%%%%%%%%%%
\begin{eqnarray}
	(\mathcal{X}^{AB})_{\alpha p_1}(\mathcal{Z}^{AB})^{\alpha p_2}&=& \,\, (\mathbf{X}^{AB}\cdot \mathbf{p}_1)(\mathbf{Z}^{AB}\cdot \mathbf{p}_2)-E_1 E_2\,(\mathbf{X}^{AB}\cdot \mathbf{Z}^{AB})+E_1(\mathbf{X}^{AB}\times \mathbf{z}^{AB})\cdot\mathbf{p}_2\nonumber\\
	&&+E_2(\mathbf{Z}^{AB}\times \mathbf{x}^{AB})\cdot\mathbf{p}_1-(\mathbf{x}^{AB}\times \mathbf{p}_1)\cdot (\mathbf{z}^{AB}\times \mathbf{p}_2) ,\qquad\qquad\qquad\quad
\end{eqnarray}
%%%%%%%%%%%%%%%%%%%%%%%%%%%%%%%%%%%%%%%%%%
%%%%%%%%%%%%% Type 3 %%%%%%%%%%%%%%%%%%%%%
%%%%%%%%%%%%%%%%%%%%%%%%%%%%%%%%%%%%%%%%%%
\begin{eqnarray}
	(\mathcal{X}^{AB})_{p_1 p_2}&=&-E_1 (\mathbf{X}^{AB}\cdot \mathbf{p}_2)+E_2 (\mathbf{X}^{AB}\cdot \mathbf{p}_1)+(\mathbf{p}_1 \times \mathbf{p}_2)\cdot \mathbf{x}^{AB} ,
\end{eqnarray}
%%%%%%%%%%%%%%%%%%%%%%%%%%%%%%%%%%%%%%%%%%
%%%%%%%%%%%%% Type 5 %%%%%%%%%%%%%%%%%%%%%
%%%%%%%%%%%%%%%%%%%%%%%%%%%%%%%%%%%%%%%%%%
\begin{eqnarray}
	(\mathcal{X}^{AB})_{\alpha \lambda}(\mathcal{Z}^{AB})_{\beta}{}^{\lambda}\epsilon^{\alpha\beta p_1 p_2}&=&(\mathbf{X}^{AB}\times\mathbf{z}^{AB})\cdot(\mathbf{p}_1\times \mathbf{p}_2)-(\mathbf{Z}^{AB}\times\mathbf{x}^{AB})\cdot(\mathbf{p}_1\times \mathbf{p}_2)\nonumber\\
	&&+E_2(\mathbf{X}^{AB}\times\mathbf{Z}^{AB})\cdot\mathbf{p}_1-E_1(\mathbf{X}^{AB}\times\mathbf{Z}^{AB})\cdot\mathbf{p}_2\quad\quad\nonumber\\
	&&-E_2(\mathbf{x}^{AB}\times\mathbf{z}^{AB})\cdot\mathbf{p}_1+E_1(\mathbf{x}^{AB}\times\mathbf{z}^{AB})\cdot\mathbf{p}_2 ,\quad\quad
\end{eqnarray}
%%%%%%%%%%%%%%%%%%%%%%%%%%%%%%%%%%%%%%%%%%
%%%%%%%%%%%%% Type 4 %%%%%%%%%%%%%%%%%%%%%
%%%%%%%%%%%%%%%%%%%%%%%%%%%%%%%%%%%%%%%%%%
\begin{eqnarray}
	(\mathcal{X}^{AB})_{\alpha p_1}(\mathcal{Z}^{AB})_{\beta p_2}\epsilon^{\alpha\beta p_3 p_4}\,\,\,=\,\,\,-(\mathbf{X}^{AB}\cdot \mathbf{p}_1)\big[ E_2\, \mathbf{Z}^{AB}\cdot(\mathbf{p}_3\times \mathbf{p}_4)-(\mathbf{z}^{AB}\times \mathbf{p}_2)\cdot(\mathbf{p}_3\times \mathbf{p}_4) \big]\, \nonumber\\
	+(\mathbf{Z}^{AB}\cdot \mathbf{p}_2)\big[ E_1\, \mathbf{X}^{AB}\cdot(\mathbf{p}_3\times \mathbf{p}_4)-(\mathbf{x}^{AB}\times \mathbf{p}_1)\cdot(\mathbf{p}_3\times \mathbf{p}_4) \big]\nonumber\\
	-E_1 E_2 E_3 (\mathbf{X}^{AB}\times\mathbf{Z}^{AB})\cdot \mathbf{p}_4+E_1 E_3\big[\mathbf{X}^{AB}\times(\mathbf{z}^{AB}\times\mathbf{p}_2)\big]\cdot \mathbf{p}_4\,\,\nonumber\\
	+E_1 E_2 E_4 (\mathbf{X}^{AB}\times\mathbf{Z}^{AB})\cdot \mathbf{p}_3-E_1 E_4\big[\mathbf{X}^{AB}\times(\mathbf{z}^{AB}\times\mathbf{p}_2)\big]\cdot \mathbf{p}_3\,\,\nonumber\\
	-E_2 E_3\big[\mathbf{Z}^{AB}\times(\mathbf{x}^{AB}\times\mathbf{p}_1)\big]\cdot \mathbf{p}_4-E_3\big[(\mathbf{x}^{AB}\times\mathbf{p}_1) \times (\mathbf{z}^{AB}\times\mathbf{p}_2)\big]\cdot \mathbf{p}_4\,\, \nonumber\\
	+E_2 E_4\big[\mathbf{Z}^{AB}\times(\mathbf{x}^{AB}\times\mathbf{p}_1)\big]\cdot \mathbf{p}_3+E_4\big[(\mathbf{x}^{AB}\times\mathbf{p}_1) \times (\mathbf{z}^{AB}\times\mathbf{p}_2)\big]\cdot \mathbf{p}_3 ,
\end{eqnarray}
%%%%%%%%%%%%%%%%%%%%%%%%%%%%%%%%%%%%%%%%%%
%%%%%%%%%%%%% Type 6 %%%%%%%%%%%%%%%%%%%%%
%%%%%%%%%%%%%%%%%%%%%%%%%%%%%%%%%%%%%%%%%%
\begin{eqnarray}
	(\mathcal{X}^{AB})_{\alpha p_1}\epsilon^{\alpha p_2 p_3 p_4}&=&(\mathbf{X}^{AB}\cdot \mathbf{p}_1)(\mathbf{p}_2\times \mathbf{p}_3)\cdot\mathbf{p}_4 - E_1 E_2\, \mathbf{X}^{AB}\cdot(\mathbf{p}_3\times \mathbf{p}_4)+ E_1 E_3\, \mathbf{X}^{AB}\cdot(\mathbf{p}_2\times \mathbf{p}_4)\nonumber\\
	&&- E_1 E_4\, \mathbf{X}^{AB}\cdot(\mathbf{p}_2\times \mathbf{p}_3)+E_2\, (\mathbf{x}^{AB}\times\mathbf{p}_1)\cdot(\mathbf{p}_3\times \mathbf{p}_4)\qquad\qquad\qquad\nonumber\\
	&&-E_3\, (\mathbf{x}^{AB}\times\mathbf{p}_1)\cdot(\mathbf{p}_2\times \mathbf{p}_4)+E_4\, (\mathbf{x}^{AB}\times\mathbf{p}_1)\cdot(\mathbf{p}_2\times \mathbf{p}_3),
\end{eqnarray}
%%%%%%%%%%%%%%%%%%%%%%%%%%%%%%%%%%%%%%%%%%
%%%%%%%%%%%%% Type 7 %%%%%%%%%%%%%%%%%%%%%
%%%%%%%%%%%%%%%%%%%%%%%%%%%%%%%%%%%%%%%%%%
\begin{eqnarray}
	(\mathcal{X}^{AB})_{\alpha \beta} (\mathcal{Z}^{AB})^{\beta p_1}\epsilon^{\alpha p_2 p_3 p_4}&=&-E_1\, (\mathbf{X}^{AB}\cdot\mathbf{Z}^{AB})(\mathbf{p}_2\times \mathbf{p}_3)\cdot \mathbf{p}_4+\mathbf{X}^{AB}\cdot(\mathbf{z}^{AB}\times\mathbf{p}_1)(\mathbf{p}_2\times \mathbf{p}_3)\cdot \mathbf{p}_4\nonumber\\
	&&+\,\mathbf{X}^{AB}\cdot\big[E_2 (\mathbf{p}_3\times \mathbf{p}_4)-E_3\, (\mathbf{p}_2\times \mathbf{p}_4)+E_4\, (\mathbf{p}_2\times \mathbf{p}_3)\big](\mathbf{Z}^{AB}\cdot \mathbf{p}_1)\nonumber\\
	&&-(\mathbf{x}^{AB}\times\mathbf{Z}^{AB})\cdot\big[E_1 E_2(\mathbf{p}_3\times \mathbf{p}_4)-E_1 E_3(\mathbf{p}_2\times \mathbf{p}_4)+E_1 E_4(\mathbf{p}_2\times \mathbf{p}_3)\big]\nonumber\\
	&&\big[\mathbf{x}^{AB}\times (\mathbf{z}^{AB}\times\mathbf{p}_1)\big]\cdot\big[E_2(\mathbf{p}_3\times \mathbf{p}_4)-E_3(\mathbf{p}_2\times \mathbf{p}_4)+E_4(\mathbf{p}_2\times \mathbf{p}_3)\big] ,\nonumber\\
\end{eqnarray}
where $p_r$ is a four momentum with components $(p_r)^{\mu}=(E_r,\mathbf{p}_r)$, with $r=1,2,3,4$. The usual three-vector products are defined by
\begin{eqnarray}
	\mathbf{a}\cdot\mathbf{b}&=&	\mathbf{a}^{i}\, \mathbf{b}^{i},\\
	(\mathbf{a}\times\mathbf{b})^{i}&=&\epsilon^{ijk}\mathbf{a}^{j}\, \mathbf{b}^{k}. \label{A10}
\end{eqnarray}
For example, taking $(\mathcal{X}^{AB})_{\mu\nu}=(Y_f)^{AB}_{\mu\nu}$ and $(\mathcal{Z}^{AB})_{\mu\nu}=(Y_f)^{*BA}_{\mu\nu}$, it follows that
\begin{eqnarray}\label{A11}
	(Y_f)^{AB}_{\mu\nu}(Y_f)^{*BA\mu\nu}&=&-2 \, \mathbf{e}^{AB}_{f}\cdot \mathbf{e}^{BA*}_{f}+2\, \mathbf{b}^{AB}_{f}\cdot \mathbf{b}^{BA*}_{f} .
\end{eqnarray}

\section{Lorentz tensors}\label{AppendixLorentzStructures}
In this appendix, we present the various Lorentz tensors that arise from the second-order contributions to the $f_B f_A \gamma$ vertex given in Eq.~(\ref{SeconOrderVertex}).
\begin{eqnarray}
	\frac{(T^{AB}_{V})_{\mu}}{c_{\Delta}} & = & \,\,\, \, \left(m_C-m_A\right) m_{AC}^2 \left(\mathcal{A}_f^{A
		C}\right)_{\mu \lambda } \left(\mathcal{A}_f^{C
		B}\right)^{\lambda P}+\left(m_B-m_C\right)
	m_{BC}^2 \left(\mathcal{A}_f^{A C}\right)^{\lambda P}
	\left(\mathcal{A}_f^{C B}\right)_{\mu \lambda}\nonumber\\
	&&+\left(m_A+m_C\right) m_{AC}^2 \left(\mathcal{V}_f^{A
		C}\right)_{\mu \lambda } \left(\mathcal{V}_f^{C
		B}\right)^{\lambda P}-\left(m_B+m_C\right)
	m_{BC}^2 \left(\mathcal{V}_f^{A C}\right)^{\lambda P}
	\left(\mathcal{V}_f^{C B}\right)_{\mu \lambda}\nonumber\\
	&&+\left(m_C-m_A\right) m_{BA}^2 \left(\mathcal{A}_f^{A
		C}\right)_{\mu \lambda } \left(\mathcal{A}_f^{C
		B}\right)^{\lambda q}+\left(m_B-m_C\right)
	m_{BA}^2 \left(\mathcal{A}_f^{A C}\right)^{\lambda q}
	\left(\mathcal{A}_f^{C B}\right)_{\mu \lambda}\nonumber \\
	&& +\left(m_A+m_C\right) m_{BA}^2 \left(\mathcal{V}_f^{A
		C}\right)_{\mu \lambda } \left(\mathcal{V}_f^{C
		B}\right)^{\lambda q}-\left(m_B+m_C\right)
	m_{BA}^2 \left(\mathcal{V}_f^{A C}\right)^{\lambda q}
	\left(\mathcal{V}_f^{C B}\right)_{\mu \lambda}\nonumber\\
	&&+i\Big[ \left(m_B+m_C\right) m_{AC}^2 \left(\mathcal{A}_f^{A
		C}\right)^{\lambda q} \big(\tilde{\mathcal{V}_f}^{C
		B}\big)_{\mu \lambda }+\left(m_A+m_C\right)
	m_{BC}^2 \big(\tilde{\mathcal{V}_f}^{A C}\big)_{\mu \lambda}\left(\mathcal{A}_f^{C B}\right)^{\lambda q}\nonumber\\
	&&+\left(m_C-m_B\right) m_{AC}^2 \left(\mathcal{V}_f^{A
		C}\right)^{\lambda q} \big(\tilde{\mathcal{A}_f}^{C
		B}\big)_{\mu \lambda }+\left(m_C-m_A\right)
	m_{BC}^2 \big(\tilde{\mathcal{A}_f}^{A C}\big)_{\mu \lambda}\left(\mathcal{V}_f^{C B}\right)^{\lambda q}
	\Big],
\end{eqnarray}
%%%%%%%%%%%%%%%%%%%%%%%%%%%%%%%%%%
\begin{eqnarray}
	\frac{(T^{AB}_{A})_{\mu}}{c_{\Delta}} & = &\left(m_A-m_C\right)
	m_{AC}^2 \left(\mathcal{A}_f^{A C}\right)_{\mu \lambda }
	\left(\mathcal{V}_f^{C B}\right)^{\lambda P}-\left(m_A+m_C\right) m_{AC}^2 \left(\mathcal{V}_f^{A C}\right)_{\mu \lambda } \left(\mathcal{A}_f^{C B}\right)^{\lambda P}\nonumber\\
	&&	\left(m_C-m_B\right) m_{BC}^2 \left(\mathcal{V}_f^{A
		C}\right)^{\lambda P} \left(\mathcal{A}_f^{C B}\right)_{\mu
		\lambda }+\left(m_B+m_C\right) m_{BC}^2 \left(\mathcal{A}_f^{A
		C}\right)^{\lambda P} \left(\mathcal{V}_f^{C B}\right)_{\mu
		\lambda }\nonumber\\
	&&+\left(m_C-m_B\right) m_{BA}^2 \left(\mathcal{V}_f^{A
		C}\right)^{\lambda q} \left(\mathcal{A}_f^{C B}\right)_{\mu
		\lambda }-\left(m_A+m_C\right) m_{BA}^2 \left(\mathcal{V}_f^{A
		C}\right)_{\mu \lambda } \left(\mathcal{A}_f^{C
		B}\right)^{\lambda q}\nonumber\\
	&&+\left(m_A-m_C\right)
	m_{BA}^2 \left(\mathcal{A}_f^{A C}\right)_{\mu \lambda }
	\left(\mathcal{V}_f^{C B}\right)^{\lambda
		q}+\left(m_B+m_C\right) m_{BA}^2 \left(\mathcal{A}_f^{A
		C}\right)^{\lambda q} \left(\mathcal{V}_f^{C B}\right)_{\mu
		\lambda }\nonumber\\
	&&+i\Big[\left(m_B-m_C\right) m_{AC}^2 \left(\mathcal{A}_f^{A
		C}\right)^{\lambda q} \big(\tilde{\mathcal{A}_f}^{C
		B}\big)_{\mu \lambda }+\left(m_A-m_C\right)
	m_{BC}^2 \big(\tilde{\mathcal{A}_f}^{A C}\big)_{\mu \lambda}\left(\mathcal{A}_f^{C B}\right)^{\lambda q}
	\nonumber\\
	&&-\left(m_B+m_C\right) m_{AC}^2 \left(\mathcal{V}_f^{A
		C}\right)^{\lambda q} \big(\tilde{\mathcal{V}_f}^{C
		B}\big)_{\mu \lambda }-\left(m_A+m_C\right)
	m_{BC}^2 \big(\tilde{\mathcal{V}_f}^{A C}\big)_{\mu \lambda }\left(\mathcal{V}_f^{C B}\right)^{\lambda q}
	\Big] ,
\end{eqnarray}
%%%%%%%%%%%%%%%%%%%%%%%%%%%%%%%%%%
\begin{eqnarray}
	\frac{(F^{AB}_{M})^{\lambda\rho}}{c_{\Delta} (m_A+m_B)}\,\,= \,\,\,  \left(m_A-m_C\right) m_{BC}^2 \left(\mathcal{A}_f^{A
		C}\right)^{\omega \lambda} \left(\mathcal{A}_f^{C
		B}\right)_{\omega}{}^{\rho}-\left(m_B-m_C\right)
	m_{AC}^2 \left(\mathcal{A}_f^{A C}\right)^{\omega \rho}
	\left(\mathcal{A}_f^{C B}\right)_{\omega}{}^{\lambda}\nonumber\\
	-\left(m_A+m_C\right)
	m_{BC}^2 \left(\mathcal{V}_f^{A C}\right)^{\omega \lambda}
	\left(\mathcal{V}_f^{C B}\right)_{\omega}{}^{\rho}+\left(m_B+m_C\right) m_{AC}^2 \left(\mathcal{V}_f^{A
		C}\right)^{\omega \rho } \left(\mathcal{V}_f^{C
		B}\right)_{\omega}{}^{\lambda} ,
\end{eqnarray}
%%%%%%%%%%%%%%%%%%%%%%%%%%%%%%%%%%
\begin{eqnarray}
	\frac{(F^{AB}_{E})^{\lambda\rho}}{c_{\Delta} (m_B-m_A)}\,\,= \,\, -\left(m_A-m_C\right) m_{BC}^2 \left(\mathcal{A}_f^{A
		C}\right)^{\omega \lambda } \left(\mathcal{V}_f^{C
		B}\right)_{\omega}{}^{\rho}-\left(m_B+m_C\right)
	m_{AC}^2 \left(\mathcal{A}_f^{A C}\right)^{\omega \rho }
	\left(\mathcal{V}_f^{C B}\right)_{\omega}{}^{\lambda}\nonumber\\
	+\left(m_A+m_C\right)
	m_{BC}^2 \left(\mathcal{V}_f^{A C}\right)^{\omega \lambda}
	\left(\mathcal{A}_f^{C B}\right)_{\omega}{}^{\rho}	+\left(m_B-m_C\right) m_{AC}^2 \left(\mathcal{V}_f^{A
		C}\right)^{\omega \rho} \left(\mathcal{A}_f^{C
		B}\right)_{\omega}{}^{\lambda} ,
\end{eqnarray}
%%%%%%%%%%%%%%%%%%%%%%%%%%%%%%%%%%
\begin{eqnarray}
	\frac{(R^{AB}_V)^{\rho\tau}}{c_{\Delta}}&=&m_{BC}^2 \left(\mathcal{A}_f^{A C}\right)^{\rho P} \left(\mathcal{V}_f^{C
		B}\right)^{\tau q}+m_{AC}^2 \left(\mathcal{A}_f^{A
		C}\right)^{\rho q} \left(\mathcal{V}_f^{C B}\right)^{\tau
		P}+m_{BA}^2 \left(\mathcal{A}_f^{A C}\right)^{\rho q}
	\left(\mathcal{V}_f^{C B}\right)^{\tau q}\nonumber\\
	&&-(\mathcal{V}_f\leftrightarrow \mathcal{A}_f) ,
\end{eqnarray}
%%%%%%%%%%%%%%%%%%%%%%%%%%%%%%%%%%
\begin{eqnarray}
	\frac{(R^{AB}_A)^{\rho\tau}}{c_{\Delta}}&=&m_{BC}^2 \left(\mathcal{V}_f^{A C}\right)^{\rho P} \left(\mathcal{V}_f^{C
		B}\right)^{\tau q}+m_{AC}^2 \left(\mathcal{V}_f^{A
		C}\right)^{\rho q} \left(\mathcal{V}_f^{C B}\right)^{\tau
		P}+m_{BA}^2 \left(\mathcal{V}_f^{A C}\right)^{\rho q}
	\left(\mathcal{V}_f^{C B}\right)^{\tau q}\nonumber\\
	&&-(\mathcal{V}_f\rightarrow \mathcal{A}_f) ,
\end{eqnarray}
%%%%%%%%%%%%%%%%%%%%%%%%%%%%%%%%%%
\begin{eqnarray}
	\frac{(S^{AB}_V)_{\mu\lambda}}{c_{\Delta}}=m_{AC}^2 \Big[\left(\mathcal{V}_f^{A
		C}\right)_{\lambda q} \left(\mathcal{V}_f^{B
		C}\right)_{\mu q}-\left(\mathcal{V}_f^{A C}\right)_{\mu P} \left(\mathcal{V}_f^{B
		C}\right)_{\lambda P}+\left(\mathcal{A}_f^{A C}\right)_{\mu P} \left(\mathcal{A}_f^{B
		C}\right)_{\lambda P}-\left(\mathcal{A}_f^{A
		C}\right)_{\lambda q} \left(\mathcal{A}_f^{B
		C}\right)_{\mu q} \nonumber \\
	-\left(m_A+m_B\right)
	\left(m_B+m_C\right) \left(\mathcal{V}_f^{A C}\right)_{\mu
		\rho} \left(\mathcal{V}_f^{C B}\right)^{\rho}{}_{\lambda}+\left(m_A+m_B\right)
	\left(m_B-m_C\right) \left(\mathcal{A}_f^{A C}\right)_{\mu
		\rho} \left(\mathcal{A}_f^{C B}\right)^{\rho}{}_{\lambda}\Big]\nonumber\\
	+m_{BC}^2
	\Big[\left(\mathcal{V}_f^{A
		C}\right)_{\lambda P} \left(\mathcal{V}_f^{B
		C}\right)_{\mu P}-\left(\mathcal{V}_f^{A
		C}\right)_{\mu q} \left(\mathcal{V}_f^{B
		C}\right)_{\lambda q} +\left(\mathcal{A}_f^{A
		C}\right)_{\mu q} \left(\mathcal{A}_f^{B
		C}\right)_{\lambda q}-\left(\mathcal{A}_f^{A
		C}\right)_{\lambda P} \left(\mathcal{A}_f^{B
		C}\right)_{\mu P}\nonumber\\
	+\left(m_A+m_B\right) \left(m_A+m_C\right)
	\left(\mathcal{V}_f^{A C}\right)^{\rho}{}_{\lambda }
	\left(\mathcal{V}_f^{C B}\right)_{\mu \rho }-\left(m_A+m_B\right) \left(m_A-m_C\right)
	\left(\mathcal{A}_f^{A C}\right)^{\rho}{}_{\lambda}
	\left(\mathcal{A}_f^{C B}\right)_{\mu \rho}\Big]\nonumber\\
	+m_{BA}^2
	\Big[\left(\mathcal{V}_f^{A C}\right)_{\lambda q}
	\left(\mathcal{V}_f^{C B}\right)_{\mu P}-\left(\mathcal{V}_f^{A
		C}\right)_{\mu P} \left(\mathcal{V}_f^{B
		C}\right)_{\lambda q}+\left(\mathcal{A}_f^{A C}\right)_{\mu P} \left(\mathcal{A}_f^{B
		C}\right)_{\lambda q}-\left(\mathcal{A}_f^{A
		C}\right)_{\lambda q} \left(\mathcal{A}_f^{B
		C}\right)_{\mu P}\Big]\nonumber\\
	+i(m_B-m_A)\Big\{ m_{AC}^2\Big[\left(m_B+m_C\right) \left(\mathcal{A}_f^{A
		C}\right)_{\mu \rho} \big(\tilde{\mathcal{V}_f}^{B
		C}\big)^{\rho}{}_{\lambda}+\left(m_C-m_B\right)
	\left(\mathcal{V}_f^{A C}\right)_{\mu \rho}
	\big(\tilde{\mathcal{A}_f}^{C B}\big)^{\rho}{}_{\lambda} \Big]  , \nonumber\\
	+m_{BC}^2\Big[\left(m_A+m_C\right) \left(\tilde{\mathcal{V}_f}^{A
		C}\right)^{\rho}{}_{\lambda} \big(\mathcal{A}_f^{B
		C}\big)_{\mu \rho}+\left(m_C-m_A\right)
	\left(\tilde{\mathcal{A}_f}^{A C}\right)^{\rho}{}_{\lambda}
	\big(\mathcal{V}_f^{C B}\big)_{\mu \rho} \Big]\Big\} , \nonumber\\
\end{eqnarray}
\begin{eqnarray}
	\frac{(S^{AB}_A)_{\mu\lambda}}{c_{\Delta}}=m_{AC}^2 \Big[\left(\mathcal{V}_f^{A
		C}\right)_{\mu P} \left(\mathcal{A}_f^{B
		C}\right)_{\lambda P}-\left(\mathcal{A}_f^{A
		C}\right)_{\mu P} \left(\mathcal{V}_f^{B
		C}\right)_{\lambda P}-\left(\mathcal{V}_f^{A
		C}\right)_{\lambda q} \left(\mathcal{A}_f^{B
		C}\right)_{\mu q}+\left(\mathcal{A}_f^{A
		C}\right)_{\lambda q} \left(\mathcal{V}_f^{B
		C}\right)_{\mu q}\nonumber\\
	+\left(m_B-m_A\right)
	\left(m_B-m_C\right) \left(\mathcal{V}_f^{A C}\right)_{\mu
		\rho } \left(\mathcal{A}_f^{C B}\right)^{\rho}{}_{\lambda}+\left(m_A-m_B\right) \left(m_B+m_C\right)
	\left(\mathcal{A}_f^{A C}\right)_{\mu \rho } \left(\mathcal{V}_f^{B
		C}\right)^{\rho}{}_{\lambda}\Big]\nonumber\\
	+m_{BC}^2\Big[ \left(\mathcal{A}_f^{A
		C}\right)_{\lambda P} \left(\mathcal{V}_f^{C B}\right)_{\mu
		P}-\left(\mathcal{V}_f^{A C}\right)_{\lambda P} \left(\mathcal{A}_f^{B
		C}\right)_{\mu P}+\left(\mathcal{V}_f^{A C}\right)_{\mu q} \left(\mathcal{A}_f^{B
		C}\right)_{\lambda q}-\left(\mathcal{A}_f^{A C}\right)_{\mu
		q} \left(\mathcal{V}_f^{C B}\right)_{\lambda q}\nonumber\\
	+\left(m_B-m_A\right) \left(m_A+m_C\right) \left(\mathcal{V}_f^{A
		C}\right)^{\rho}{}_{\lambda } \left(\mathcal{A}_f^{B
		C}\right)_{\mu \rho }+\left(m_A-m_B\right)
	\left(m_A-m_C\right) \left(\mathcal{A}_f^{A C}\right)^{\rho}{}_{\lambda} \left(\mathcal{V}_f^{C B}\right)_{\mu \rho }\Big]\nonumber\\
	+m_{BA}^2\Big[\left(\mathcal{V}_f^{A C}\right)_{\mu
		P} \left(\mathcal{A}_f^{C B}\right)_{\lambda q}-\left(\mathcal{V}_f^{A C}\right)_{\lambda q} \left(\mathcal{A}_f^{B
		C}\right)_{\mu P}+\left(\mathcal{A}_f^{A C}\right)_{\lambda q} \left(\mathcal{V}_f^{B
		C}\right)_{\mu P}-\left(\mathcal{A}_f^{A
		C}\right)_{\mu P} \left(\mathcal{V}_f^{C B}\right)_{\lambda
		q} \Big]\nonumber\\
	-i(m_A+m_B)\Big\{ m_{AC}^2\Big[\left(m_B-m_C\right) \left(\mathcal{A}_f^{A
		C}\right)_{\mu \rho} \big(\tilde{\mathcal{A}_f}^{B
		C}\big)^{\rho}{}_{\lambda}-\left(m_C+m_B\right)
	\left(\mathcal{V}_f^{A C}\right)_{\mu \rho}
	\big(\tilde{\mathcal{V}_f}^{C B}\big)^{\rho}{}_{\lambda} \Big]  , \nonumber\\
	+m_{BC}^2\Big[\left(m_A-m_C\right) \left(\tilde{\mathcal{A}_f}^{A
		C}\right)^{\rho}{}_{\lambda} \big(\mathcal{A}_f^{B
		C}\big)_{\mu \rho}-\left(m_A+m_C\right)
	\left(\tilde{\mathcal{V}_f}^{A C}\right)^{\rho}{}_{\lambda}
	\big(\mathcal{V}_f^{C B}\big)_{\mu \rho} \Big]\Big\} , \nonumber\\
\end{eqnarray}

\begin{eqnarray}
	\frac{2(\hat{F}^{AB}_{M})^{\mu\lambda\rho}}{c_{\Delta}(m_A+m_B)}=m_{AC}^2 \Big[\left(m_B-m_C\right) \left(
	\mathcal{A}_f^{A C}\right)^{\mu P}
	\left(\mathcal{A}_f^{C B}\right)^{\lambda \rho }-\left(m_B+m_C\right)
	\left(\mathcal{V}_f^{A C}\right)^{\mu P} \left(\mathcal{V}_f^{C B}\right)^{\lambda \rho }\nonumber\\
	+2
	\left(m_A-m_C\right) \left(\mathcal{A}_f^{A C}\right)^{\mu \lambda} \left(\mathcal{A}_f^{C B}\right)^{\rho P}-2 \left(m_A+m_C\right)
	\left(\mathcal{V}_f^{A C}\right)^{\mu \lambda} \left(\mathcal{V}_f^{B
		C}\right)^{\rho P}\Big]\nonumber\\
	+m_{BC}^2\Big[\left(m_A-m_C\right) \left(\mathcal{A}_f^{A C}\right)^{\lambda \rho }
	\left(\mathcal{A}_f^{C B}\right)^{\mu P}-\left(m_A+m_C\right)
	\left(\mathcal{V}_f^{A C}\right)^{\lambda \rho} \left(\mathcal{V}_f^{B
		C}\right)^{\mu P}\nonumber\\
	+2 \left(m_B-m_C\right)
	\left(\mathcal{A}_f^{A C}\right)^{\rho P} \left(\mathcal{A}_f^{C B}\right)^{\mu \lambda}-2 \left(m_B+m_C\right) \left(\mathcal{V}_f^{A
		C}\right)^{\rho P} \left(\mathcal{V}_f^{C B}\right)^{\mu \lambda}\Big]\nonumber\\
	+2m_{BA}^2\Big[ \left(m_A-m_C\right) \left(\mathcal{A}_f^{A C}\right)^{\mu \lambda}
	\left(\mathcal{A}_f^{C B}\right)^{\rho q}+\left(m_B-m_C\right)
	\left(\mathcal{A}_f^{A C}\right)^{\rho q} \left(\mathcal{A}_f^{B
		C}\right)^{\mu \lambda }\nonumber\\
	-\left(m_A+m_C\right)
	\left(\mathcal{V}_f^{A C}\right)^{\mu \lambda} \left(\mathcal{V}_f^{B
		C}\right)^{\rho q}-\left(m_B+m_C\right) \left(\mathcal{V}_f^{A
		C}\right)^{\rho q} \left(\mathcal{V}_f^{C B}\right)^{\mu \lambda}\Big]\, ,
\end{eqnarray}

\begin{eqnarray}
	\frac{2(\hat{F}^{AB}_{E})^{\mu\lambda\rho}}{c_{\Delta}(m_B-m_A)}=m_{AC}^2 \Big[	\left(m_C-m_B\right) \left(\mathcal{V}_f^{A C}\right)^{\mu P}
	\left(\mathcal{A}_f^{C B}\right)^{\lambda \rho}+\left(m_B+m_C\right) \left(\mathcal{A}_f^{A
		C}\right)^{\mu P} \left(\mathcal{V}_f^{C B}\right)^{\lambda \rho}\nonumber\\
	+2\left(m_A+m_C\right) \left(\mathcal{V}_f^{A C}\right)^{\mu \lambda} \left(\mathcal{A}_f^{C B}\right)^{\rho P}+2 \left(m_C-m_A\right)
	\left(\mathcal{A}_f^{A C}\right)^{\mu \lambda} \left(\mathcal{V}_f^{B
		C}\right)^{\rho P}\Big]\nonumber\\
	+m_{BC}^2\Big[\left(m_C-m_A\right) \left(\mathcal{A}_f^{A C}\right)^{\lambda \rho}
	\left(\mathcal{V}_f^{C B}\right)^{\mu P}+\left(m_A+m_C\right)
	\left(\mathcal{V}_f^{A C}\right)^{\lambda \rho} \left(\mathcal{A}_f^{B
		C}\right)^{\mu P}\nonumber\\
	+2 \left(m_B+m_C\right)
	\left(\mathcal{A}_f^{A C}\right)^{\rho P} \left(\mathcal{V}_f^{B
		C}\right)^{\mu \lambda}+2 \left(m_C-m_B\right) \left(\mathcal{V}_f^{A
		C}\right)^{\rho P} \left(\mathcal{A}_f^{C B}\right)^{\mu \lambda}\Big]\nonumber\\
	+2m_{BA}^2\Big[ \left(m_C-m_A\right) \left(\mathcal{A}_f^{A C}\right)^{\mu \lambda}
	\left(\mathcal{V}_f^{C B}\right)^{\rho q}+\left(m_B+m_C\right)
	\left(\mathcal{A}_f^{A C}\right)^{\rho q} \left(\mathcal{V}_f^{B
		C}\right)^{\mu \lambda}\nonumber\\
	+\left(m_A+m_C\right)
	\left(\mathcal{V}_f^{A C}\right)^{\mu \lambda} \left(\mathcal{A}_f^{B
		C}\right)^{\rho q}+\left(m_C-m_B\right) \left(\mathcal{V}_f^{A
		C}\right)^{\rho q} \left(\mathcal{A}_f^{C B}\right)^{\mu \lambda}\Big] ,
\end{eqnarray}
where
\begin{eqnarray}
	c_{\Delta}=\frac{2 v^2}{m_{BA}^2 m_{AC}^2 m_{BC}^2} .
\end{eqnarray}

%%%%%%%%%%%%%%%%%%%%%%%%%%%%%%%%%%%%%%%%%%
%%%%%%%%%%%%%%%%%%%%%%%%%%%%%%%%%%%%%%%%%%
%%%%%%%%%%%%%%%%%%%%%%%%%%%%%%%%%%%%%%%%%%
\bibliography{Refs}

\end{document}